\documentclass[12pt,dviwindows]{article}

\usepackage{color,epsf,graphicx}
\usepackage{amsmath,amssymb}
\usepackage{psfrag}
\usepackage[colorlinks=true,citecolor=blue,linkcolor=red]{hyperref}

\renewcommand{\bar}[1]{\overline{#1}}

\renewcommand{\bar}[1]{\overline{#1}}

\def\ru1{\rule[-0.4truecm]{0mm}{1truecm}}

\begin{document}

\begin{flushright}

\end{flushright}


\centerline{\Large \bf Nucleon Transversity and Hyperon Polarization}

\vspace{10mm}

\centerline{{\bf Dae Sung Hwang}}

\vspace{5mm}

\vspace{4mm} \centerline{\it Institute of Fundamental Physics, Sejong
University, Seoul 05006, South Korea}

\vspace{20mm}

\centerline{\bf Abstract}

\vspace{10mm}

\noindent
We calculate the transverse polarizations of the produced hyperons in the semi-inclusive
deep inelastic scattering of unpolarized lepton beam on transversely polarized nucleon,
since these polarizations provide a potential method for extracting the transversity
distribution $h_1(x)$ of the nucleon.
In this calculation we use the SU(6) wavefunctions of the octet baryons and the spectator model
for the distribution functions of nucleons and the fragmentation functions of hyperons.
We find that that $H_1(z)$ of the $\Sigma$ hyperons are much larger than
that of the $\Lambda$ hyperon.
Therefore, when one tries to extract the transversity distribution $h_1(x)$
from the hyperon polarizations,
measuring the polarizations of the $\Sigma$ hyperons is more efficient than $\Lambda$.

\vspace{0.5cm}

\noindent \vspace*{12mm}


\vspace{3mm}


\newpage

\section{Introduction}

The unpolarized distribution $f_1(x)$ and the helicity distribution $g_1(x)$
of nucleon have been extensively investigated.
However, the transversity distribution $h_1(x)$ is less known since it can not be measured in
fully deep inelastic scattering since it is chiral-odd.
When the spins of two hadron beams in the hadron collider are transversely
polarized, $h_1(x)$ can be extracted by measuring the double-spin transverse
asymmetry \cite{BDR02}.
Besides this, an important approach is using
the Collins mechanism \cite{Collins:1992kk}
in semi-inclusive deep inelastic scattering (SIDIS).
By using this method, HERMES, COMPASS and CLAS have been
making important progresses \cite{Anselmino:2007fs}.
In applying the Collins mechanism to the SIDIS process
$lp^{\uparrow}\to l\pi X$ (where $p$ is proton or neutron),
one needs to separate the contributions from
the Collins and Sivers mechanisms \cite{Collins:1992kk,Sivers}.

In this paper we study another method which can be used in
extracting the transversity distribution $h_1(x)$,
which was presented in Refs. \cite{Kunne:1993nq,BM98,Anselmino03}.
This method considers the process $lp^{\uparrow}\to l\Lambda^{\uparrow}X$ with an unpolarized
lepton beam, a transversely polarized proton or neutron target ($S_N$) and the measurement of the
transverse polarization $P_N$ of a produced $\Lambda$ hyperon;
transverse means orthogonal to the $\gamma^* - \Lambda$ plane
\cite{Kunne:1993nq,BM98,Anselmino03}:
\begin{equation}
P_N={2(1-y)\over 1+(1-y)^2}\
{\Sigma_q e_q^2h_{1q}(x) H_{1q}(z)\over \Sigma_q e_q^2f_{1q}(x) D_{1q}(z)}\ .
\label{TP1}
\end{equation}
When we neglect the sea quark contributions (which should be safe in the
large $x$ and $z$ regions), (\ref{TP1}) becomes
\begin{equation}
P_N={2(1-y)\over 1+(1-y)^2}\
{4h_{1u}(x) H_{1u}(z) + h_{1d}(x) H_{1d}(z)\over
4f_{1u}(x) D_{1u}(z) + f_{1d}(x) D_{1d}(z)} \ .
\label{TP2}
\end{equation}

The polarizations of
$\Lambda$, $\Sigma^+$, $\Sigma^0$, $\Sigma^-$, $\Xi^0$ and $\Xi^-$
can be measured by measuring the angular distributions of the following
two-body decays \cite{Agashe:2014kda}:
\begin{eqnarray}
{\Lambda}(1116\ {\rm MeV})&\longrightarrow&p\pi^-\ (63.9\%)\ ,\ \ \
n\pi^0(35.8\%) \ ,
\label{su66}\\
\Sigma^+(1189\ {\rm MeV})&\longrightarrow&p\pi^0\ (51.6\%)\ ,\ \ \
n\pi^+(48.3\%) \ ,
\label{su67}\\
\Sigma^0(1193\ {\rm MeV})&\longrightarrow&\Lambda\gamma\ (100\%)\ ,
\label{su67a}\\
\Sigma^-(1197\ {\rm MeV})&\longrightarrow&n\pi^-\ (99.8\%)\ ,
\label{su68}\\
\Xi^0(1315\ {\rm MeV})&\longrightarrow&\Lambda\pi^0\ (99.5\%)\ ,
\label{su69}\\
\Xi^-(1322\ {\rm MeV})&\longrightarrow&\Lambda\pi^-\ (99.9\%)\ .
\label{su60}
\end{eqnarray}

The COMPASS group performed experiment
\cite{Negrini:2009zz,Negrini:2009oia,Collaboration:2011zba}
measuring the transverse polarizations of
$\Lambda$ and $\bar{\Lambda}$ produced in the SIDIS process in which
160 GeV longitudinally polarized muon beam was incident on the transversely
polarized ${\rm NH}_3$ target, and obtained the result that
the transverse polarizations of $\Lambda$ and $\bar{\Lambda}$ are compatible
with zero within their error bars and they have no dependence on $x$ or $z$.
Eq. (\ref{TP1}) represents the transfer of transverse spins from an initial
nucleon to a produced hyperon, and it is satisfied for a polarized lepton beam
as well as for an unpolarized lepton beam.
This COMPASS experiment selected events at $Q^2>1$ GeV and $0.1<y<0.9$, and
in the kinematic range $8\cdot 10^{-3}<x<0.1$ and $0.1<z<0.5$.
So, it seems that it is difficult to investigate the nucleon transversity
through the measurement of the $\Lambda$ polarization using Eq. (\ref{TP2}).
There were various speculations about the smallness of the measured transverse
polarizations of $\Lambda$ and $\bar{\Lambda}$ such as:
It could be attributed to the cancellation of the $u$ and $d$ quark
contributions \cite{Negrini:2009zz}.
It might be due to the smallness of the transversity distribution in the
available $x$-range or to the fact that the polarized fragmentation function
into a transverse lambda is small in the COMPASS kinematic range \cite{Collaboration:2011zba}.
In this paper we try to understand this question.

In this paper
we calculate the transverse polarizations of the produced hyperons in the SIDIS process
of unpolarized lepton beam on transversely polarized nucleon,
with the SU(6) wavefunctions of the octet baryons and the spectator model
for the distribution functions of nucleons and the fragmentation functions of hyperons.
Ref. \cite{Yang:2002gh} performed the similar calculation. However, since the purpose of this paper
is to understand why the COMPASS group obtained the polarization of $\Lambda$
compatible with zero and how we can overcome this difficulty in investigating the nucleon
transversity through the hyperon polarization measurement, we present the results in
detail and show clearly that the sizes of the polarizations of $\Lambda$ and $\Sigma$ are
expected to be very different, and the polarization of $\Sigma$ is much larger than
that of $\Lambda$.

We find that that $H_1(z)$ of the $\Lambda$ hyperon 
is very small for both initial $u$ and $d$ quarks, and therefore
the polarization of the produced $\Lambda$ is very small for both transversely
polarized proton and neutron targets.
On the other hand, $H_1(z)$ of $\Sigma^+$ is large for initial $u$ quark and
$H_1(z)$ of $\Sigma^0$ is large for both initial $u$ and $d$ quarks, and
$H_1(z)$ of $\Sigma^-$ is large for initial $d$ quark.
Then, we find that the polarization of the produced $\Sigma^+$ is large
for proton target, that of $\Sigma^0$ is large for both proton and neutron targets,
and that of $\Sigma^-$ large for neutron target.
Therefore, when one tries to extract the transversity distribution $h_1(x)$ of nucleon
from the hyperon polarization in the SIDIS process $lp^{\uparrow}\to lH^{\uparrow}X$
(where $p$ is proton or neutron),
measuring the polarizations of the $\Sigma^+$ and $\Sigma^0$ hyperons is more efficient
than $\Lambda$ for the proton target,
and measuring the polarizations of the $\Sigma^-$ and $\Sigma^0$ hyperons is more efficient 
than $\Lambda$ for the neutron target.
When we consider that the polarization of ${\Sigma}^0$ is measured throuth the daughter
$\Lambda$ polarization,
measuring the polarization of $\Sigma^+$ is most efficient for the proton target,
and $\Sigma^-$ for the neutron target.

\vfill

\section{SU(6) Wavefunction}

For the octet baryons, we have the following relations for fragmentation functions
from the SU(6) wavefunctions given in Appendix A \cite{VanRoyen:1967nq,Jakob:1993th}:
\begin{eqnarray}
D(u\rightarrow p)&=&2\ \Big(\ {3\over 4}D^s+{1\over 4}D^a\ \Big)\ ,
\ \ \ D(d\rightarrow p)\ =\ D^a \ .
\label{su12p}\\
D(u\rightarrow n)&=&D^a\ ,
\ \ \ D(d\rightarrow n)\ =\ 2\ \Big(\ {3\over 4}D^s+{1\over 4}D^a\ \Big) \ .
\label{su12n}\\
D(u\rightarrow \Lambda)&=&{1\over 4}D^s+{3\over 4}D^a\ ,\ \ \ 
D(d\rightarrow \Lambda)\ =\ {1\over 4}D^s+{3\over 4}D^a\ ,\ \ \
D(s\rightarrow \Lambda)\ =\ D^s \ .
\qquad\ \ \ \
\label{su11}\\
D(u\rightarrow \Sigma^+)&=&2\ \Big(\ {3\over 4}D^s+{1\over 4}D^a\ \Big)\ ,
\ \ \ D(s\rightarrow \Sigma^+)\ =\ D^a \ .
\label{su12}\\
D(u\rightarrow \Sigma^0)&=&{3\over 4}D^s+{1\over 4}D^a\ ,\ \ \
D(d\rightarrow \Sigma^0)\ =\ {3\over 4}D^s+{1\over 4}D^a\ ,\ \ \
D(s\rightarrow \Sigma^0)\ =\ D^a \ .
\qquad\ \ \ \
\label{su11s0}\\
D(d\rightarrow \Sigma^-)&=&2\ \Big(\ {3\over 4}D^s+{1\over 4}D^a\ \Big)\ ,
\ \ \ D(s\rightarrow \Sigma^-)\ =\ D^a \ .
\label{su13}\\
D(u\rightarrow \Xi^0)&=&D^a\ ,\ \ \
D(s\rightarrow \Xi^0)\ =\ 2\ \Big(\ {3\over 4}D^s+{1\over 4}D^a\ \Big) \ .
\label{su14}\\
D(d\rightarrow \Xi^-)&=&D^a\ ,\ \ \ 
D(s\rightarrow \Xi^-)\ =\ 2\ \Big(\ {3\over 4}D^s+{1\over 4}D^a\ \Big)\ .
\label{su15}
\end{eqnarray}
In the above the superscript $s$($a$) denotes that the accompanying diquark
is a scalar (axial-vector) diquark. The same relations are also satisfied for the
distribution functions.

In order to illustrate the meaning of Eqs. (\ref{su12p}) to (\ref{su15}),
let us write the relations for proton and $\Lambda$ in detail.
Eq. (\ref{su12p}) shows the following relations of distribution functions $f_1$
and fragmentation functions $D_1$ for proton:
\begin{eqnarray}
&&f_{1u}={3\over 2}f_1^s+{1\over 2}f_1^a\ ,\qquad f_{1d}=f_1^a\ ,
\label{sa11}\\
&&D_1^{u\to p}={3\over 2}D_1^s+{1\over 2}D_1^a\ ,\qquad D_1^{d\to p}=D_1^a
\ ,
\label{sa12}
\end{eqnarray}
where the subscript $u$ and $d$ denote up and down quarks, repectively, 
and Eq. (\ref{su11}) shows the following relations for $\Lambda$:
\begin{eqnarray}
&&f_{1u}={1\over 4}f_1^s+{3\over 4}f_1^a\ ,\qquad f_{1d}={1\over 4}f_1^s+{3\over 4}f_1^a\ ,
\qquad f_{1(s)}=f_1^s\ ,
\label{sa13}\\
&&D_1^{u\to \Lambda}={1\over 4}D_1^s+{3\over 4}D_1^a\ ,
\qquad D_1^{d\to \Lambda}={1\over 4}D_1^s+{3\over 4}D_1^a\ ,
\qquad D_1^{(s)\to \Lambda}=D_1^s
\ ,
\label{sa14}
\end{eqnarray}
where $(s)$ denotes the strange quark.

\section{Spectator Model}

\subsection{Distribution Functions of Nucleon}

In this paper we use the spectator model of Jakob et al. \cite{JMR97},
which takes the following baryon-quark-diquark vertex
for the scalar ($s$) and axial-vector ($a$) diquark, respectively:
\begin{equation}
\Upsilon^s={\bf{1}}g_s(p^2)\ , \qquad
\Upsilon^{a\mu}={g_a(p^2)\over {\sqrt{3}}}\gamma_5\Big( \gamma^\mu + {P^{\mu}\over M}\Big) \ ,
\label{sa1}
\end{equation}
with the following form factors:
\begin{equation}
g_{R}(p^2)=N_{R}{p^2-m^2\over |p^2-\Lambda^2|^2}\ ,
\label{sa2}
\end{equation}
where we take $\alpha =2$ for the parameter $\alpha$ in Ref. \cite{JMR97}.
In Eq. (\ref{sa1}) $P^{\mu}$ and $M$ are the momentum and mass of the baryon and
in Eq. (\ref{sa2}) $\Lambda$ is a parameter for the cut off and $p$ is the quark momentum
which satisfies 
$-p^2(x,{\bf p}_{\perp}^2)={{\bf p}_{\perp}^2\over 1-x}+{x\over 1-x}M_{R}^2-xM^2=
{{\bf p}_{\perp}^2+{\lambda}^2_R(x)\over 1-x}-{\Lambda}^2$
where ${\lambda}_{R}^2(x)$ is given by
\begin{equation}
{\lambda}_{R}^2(x) = (1-x){\Lambda}^2 + xM_R^2 - x(1-x)M^2 \ ,
\label{ssa4}
\end{equation}
in which $R=s$ or $a$.

The distribution functions $f_1$ and
$h_1$ are given by \cite{JMR97}
\begin{eqnarray}
f_1^R(x,{\bf p}_{\perp}^2)&=&
{N_R^2(1-x)^3\over 16 \pi^3}\ {(xM+m)^2+{\bf p}_{\perp}^2\over ({\bf p}_{\perp}^2+{\lambda}_{R}^2)^4}
\ ,
\label{sa4}\\
h_1^R(x,{\bf p}_{\perp}^2)&=&
a_R\ {N_R^2(1-x)^3\over 16 \pi^3}\ {(xM+m)^2\over ({\bf k}_{\perp}^2+{\lambda}_{R}^2)^4}\ ,
\label{sa5}
\end{eqnarray}
where $M$, $m$ and $M_R$ the baryon, quark and diquark mass, respectively, and
\begin{eqnarray}
f_1^R(x)&=&
{N_R^2(1-x)^3\over 48 \pi^2}\
{(xM+m)^2+{1\over 2}{\lambda}^2_{R}(x)\over \Big( {\lambda}^2_{R}(x){\Big)}^3} \ ,
\label{transv1}\\
h_1^R(x)&=&
a_R\ {N_R^2(1-x)^3\over 48 \pi^2}\
{(xM+m)^2\over \Big( {\lambda}^2_{R}(x){\Big)}^3} \ ,
\label{transv2}
\end{eqnarray}
where $a_R$ for scalar and axial-vector diquark models are given by \cite{JMR97}
\begin{equation}
a_s=1\ , \qquad a_a=-{1\over 3}\ .
\label{ssa3aa}
\end{equation}
The normalization constant $N_R$ is fixed by
\begin{equation}
\int_0^1 dx\ f_{1}^R(x)=1\ .
\label{ssa5}
\end{equation}

From the SU(6) wavefunctions given in Appendix A, we get the following relations
for proton:
\begin{equation}
f_{1u}={3\over 2}f_1^s+{1\over 2}f_1^a\ ,\qquad f_{1d}=f_1^a\ .
\label{sa11abc}
\end{equation}
and for neutron:
\begin{equation}
f_{1u}=f_1^a\ ,\qquad f_{1d}={3\over 2}f_1^s+{1\over 2}f_1^a\ .
\label{sa11abcneutron}
\end{equation}
The distribution function $h_1$ also satisfy the same relations as
the above (\ref{sa11abc}) and (\ref{sa11abcneutron}).

We use $\Lambda =0.5$, $m=0.36$, $M_s=0.6$, $M_a=0.8$ and $M=0.94$ (nucleon mass)
in the unit of GeV, when we calculate the distribution and fragmentation functions
of the proton and neutron.
We use $\Lambda =0.5$, $m=0.36$, $M_s=0.8$, $M_a=1.0$
when we calculate the fragmentation functions of $\Lambda$ and $\Sigma$ hyperons,
and $\Lambda =0.6$, $m=0.36$, $M_s=1.0$, $M_a=1.2$
when we calculate the fragmentation functions of $\Xi$ hyperons.
For the baryon mass $M$, we use 1.12 for $\Lambda$, 1.19 for $\Sigma^{\pm}$
and $\Sigma^{0}$, and 1.32 for $\Xi^0$ and $\Xi^-$.
The distribution functions of proton obtained by using the formulas in this
section are presented in Fig. 1.

\begin{figure}
\centering
\psfrag{xperp}[cc][cc]{$x_\perp$}
\begin{minipage}[t]{7.0cm}
\centering
\includegraphics[width=\textwidth]{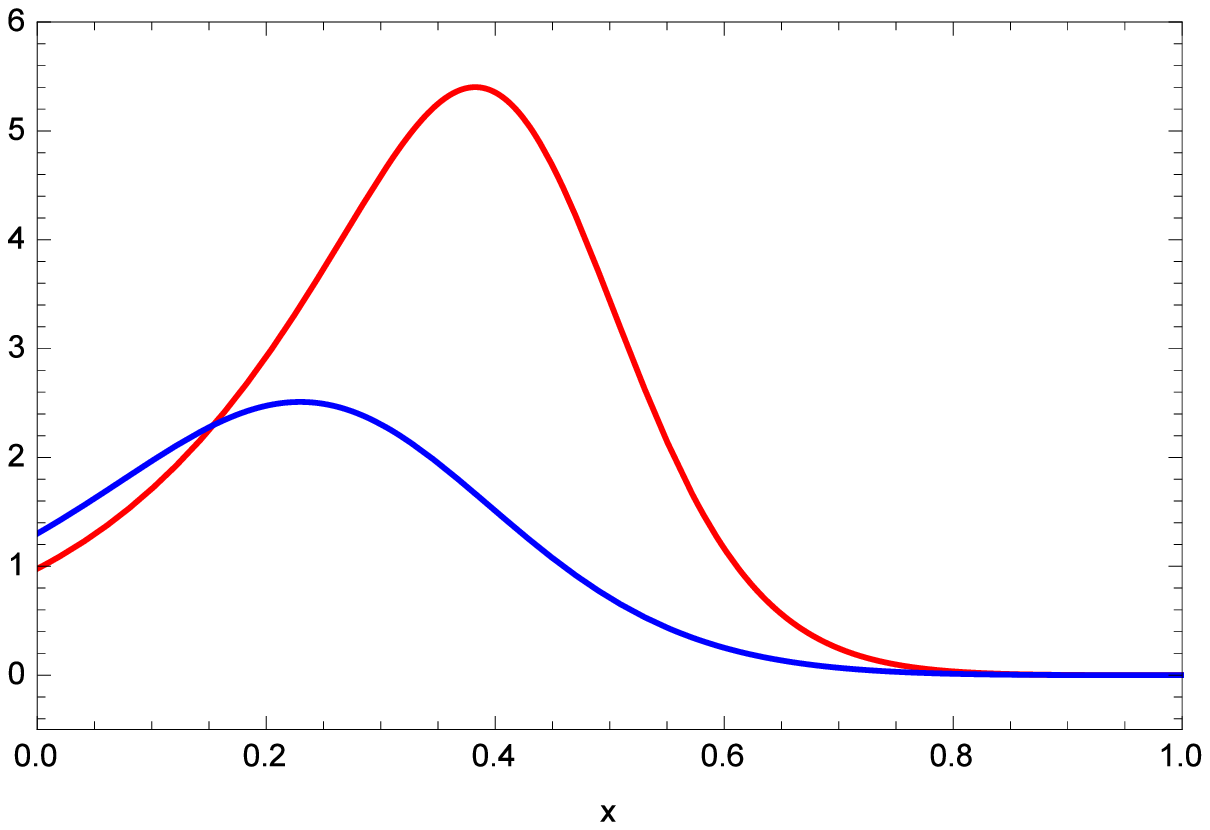}
(a)
\end{minipage}\hspace{0.5cm}
\begin{minipage}[t]{7.0cm}
\centering
\includegraphics[width=\textwidth]{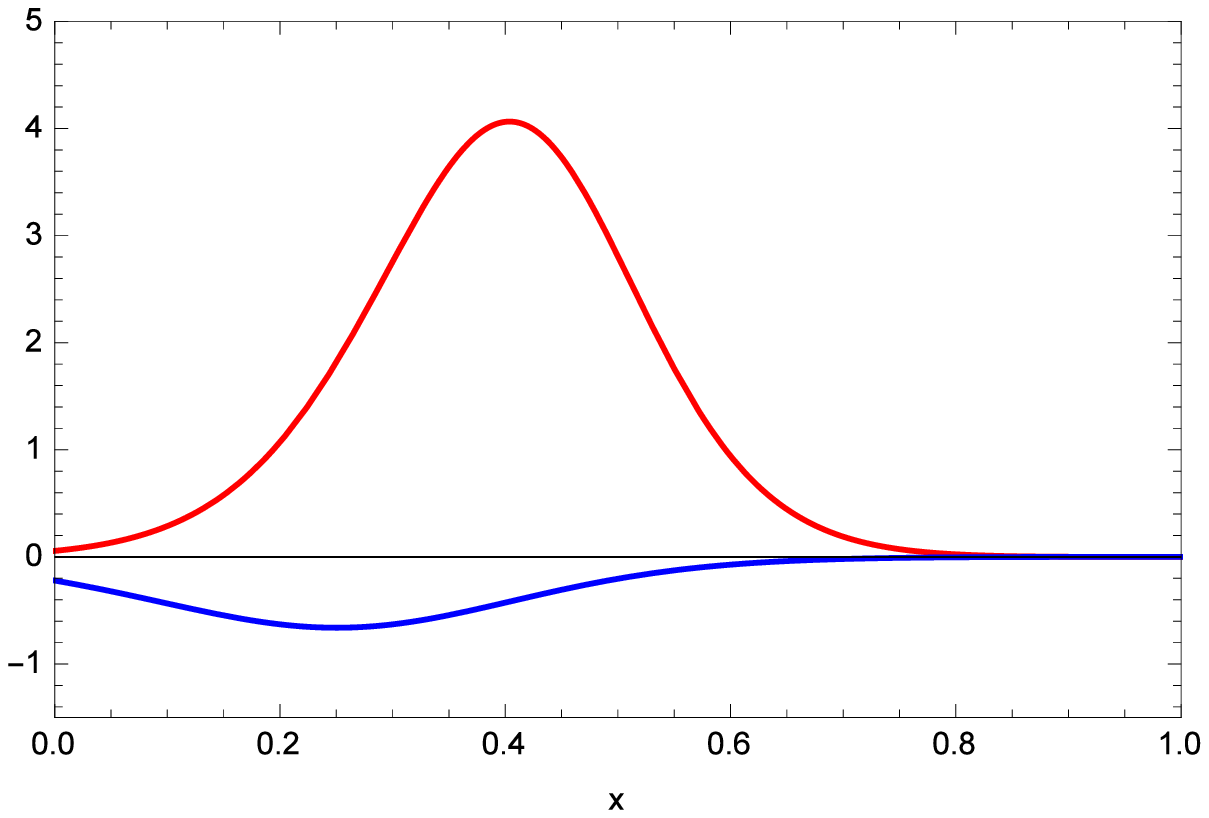}
(b)
\end{minipage}
\parbox{0.95\textwidth}{\caption{
(a) $f_{1u}(x)$ (upper line) and $f_{1d}(x)$ (lower line) of proton.
(b) $h_{1u}(x)$ (upper line) and $h_{1d}(x)$ (lower line) of proton.
(For the distribution functions of neutron,
$u$ and $d$ are exchanged in the above graphs.)
\label{f1h1udproton}}}
\end{figure}

\begin{figure}
\centering
\psfrag{xperp}[cc][cc]{$x_\perp$}
\begin{minipage}[t]{7.0cm}
\centering
\includegraphics[width=\textwidth]{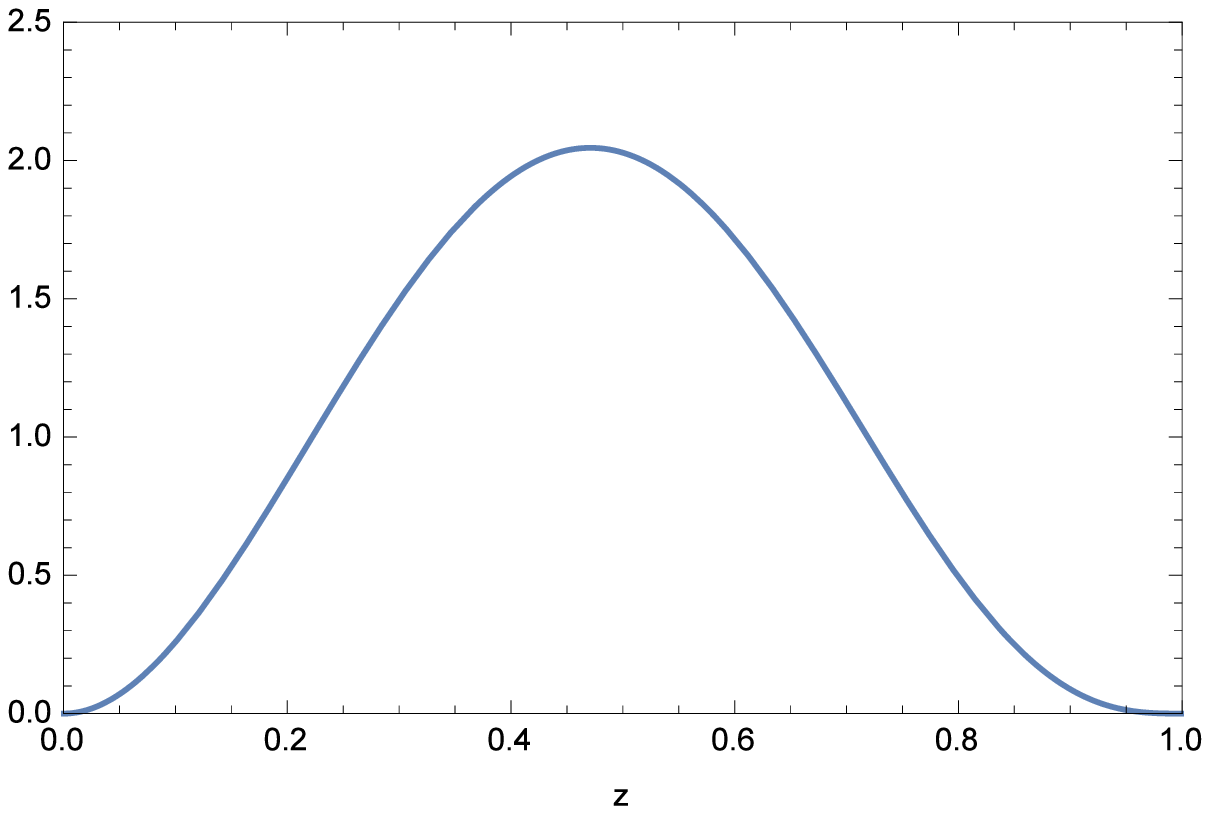}
(a)
\end{minipage}\hspace{0.5cm}
\begin{minipage}[t]{7.0cm}
\centering
\includegraphics[width=\textwidth]{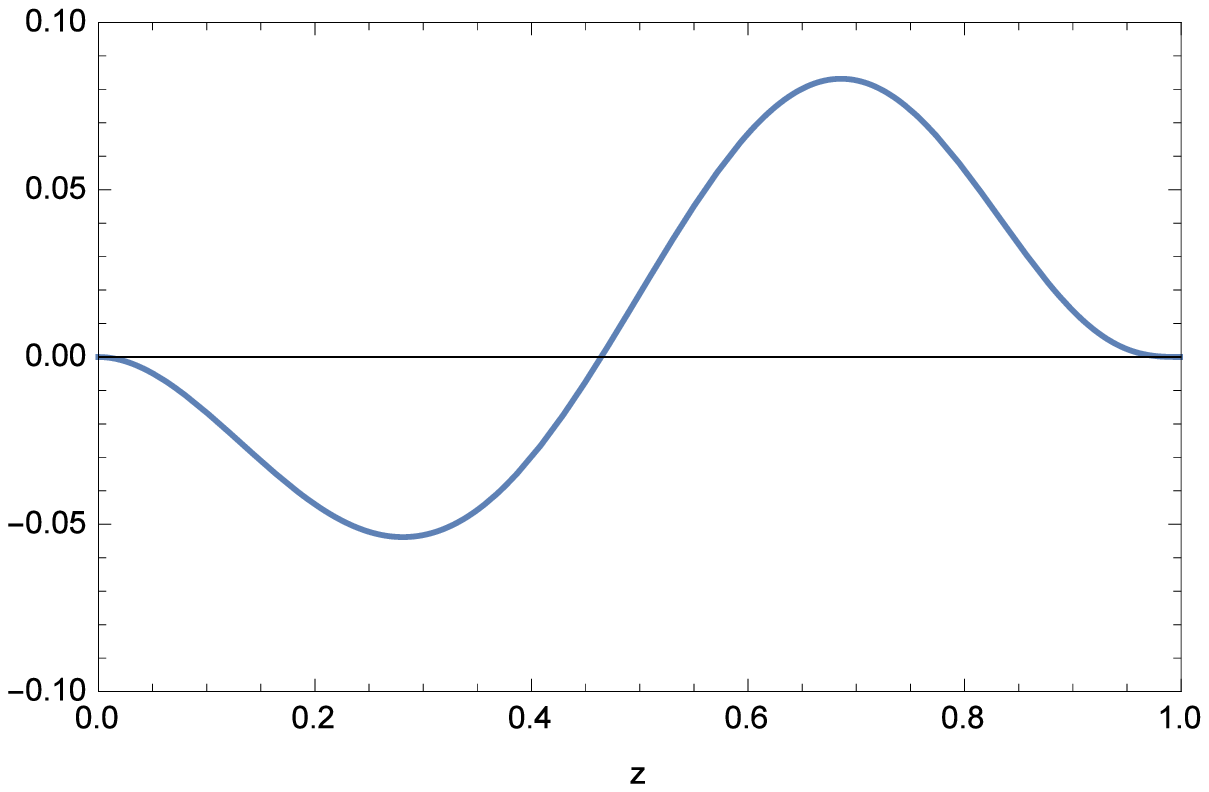}
(b)
\end{minipage}
\parbox{0.95\textwidth}{\caption{
(a) $D_1^{u\to \Lambda}(z)$ and 
(b) $H_1^{u\to \Lambda}(z)$.
We note that $D_1^{u\to \Lambda}(z)=D_1^{d\to \Lambda}(z)$ and
$H_1^{u\to \Lambda}(z)=H_1^{d\to \Lambda}(z)$.
\label{NormDH1uLambdaf}}}
\end{figure}

\begin{figure}
\centering
\psfrag{xperp}[cc][cc]{$x_\perp$}
\begin{minipage}[t]{7.0cm}
\centering
\includegraphics[width=\textwidth]{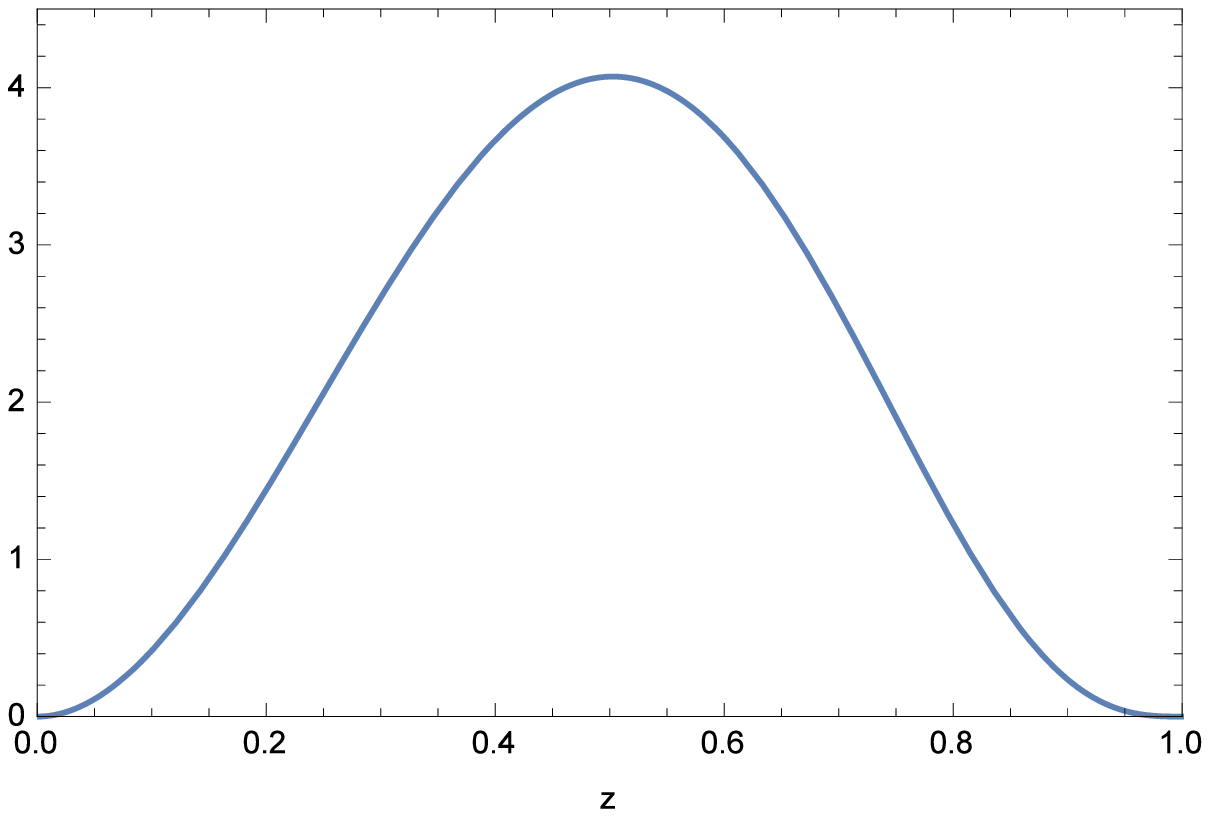}
(a)
\end{minipage}\hspace{0.5cm}
\begin{minipage}[t]{7.0cm}
\centering
\includegraphics[width=\textwidth]{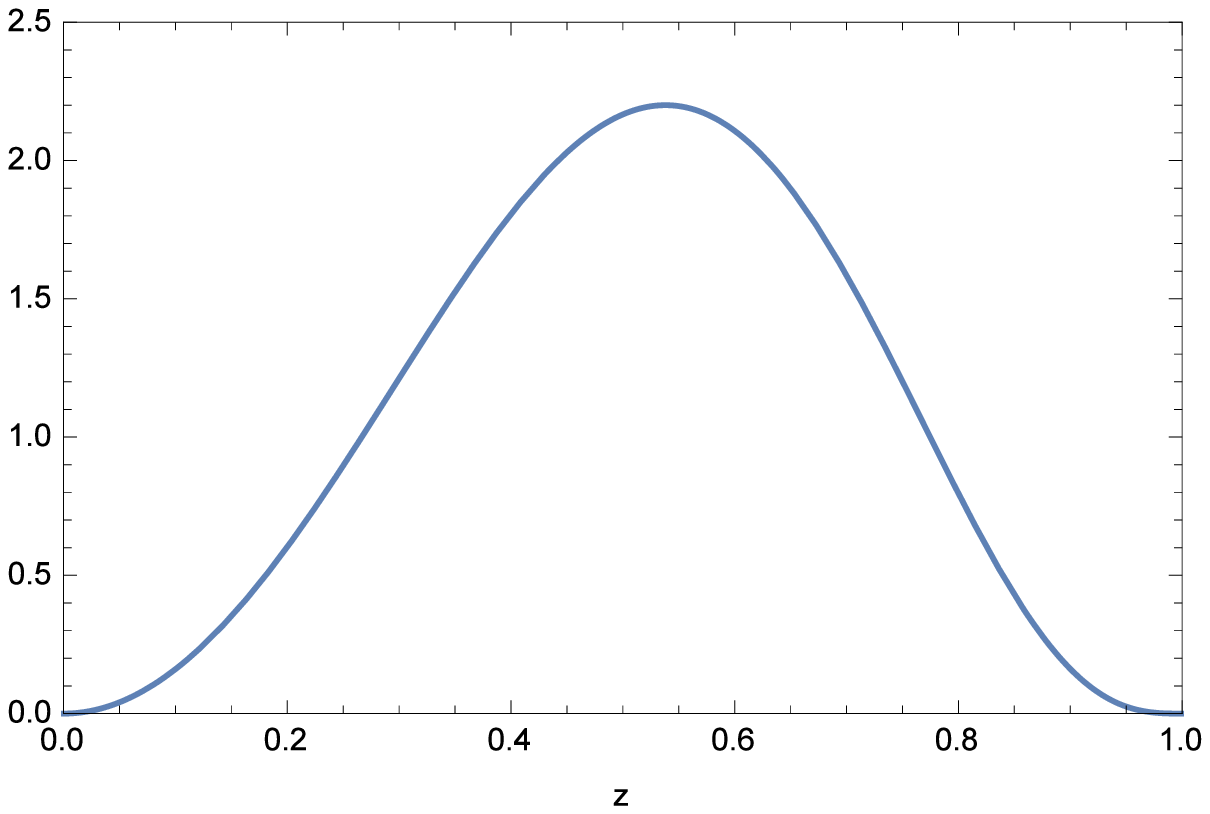}
(b)
\end{minipage}
\parbox{0.95\textwidth}{\caption{
(a) $D_1^{u\to \Sigma^+}(z)$ and 
(b) $H_1^{u\to \Sigma^+}(z)$.
\label{NormDH1uSigmaplusf}}}
\end{figure}

\subsection{Fragmentation Functions of Hyperons}

By applying to the above distribution functions $f_1$ and $h_1$
the Gribov-Lipatov reciprocity relation \cite{JMR97}
\begin{equation}
\Delta^{[\Gamma ]}(z,{\bf k}_T)={1\over 2z}\Phi^{[\Gamma' ]}({1\over z},{\bf k}_T)=
{1\over 2z}\Phi^{[\Gamma' ]}({1\over z},-{{\bf k}_T'\over z})\ ,
\label{ssa6c}
\end{equation}
the following fragmentation functions $D_1$ and $H_1$ are obtained \cite{JMR97}:
\begin{eqnarray}
D_1^R(z,z^2{\bf k}_{\perp}^2)&=&
{{\bar N}_R^2(1-z)^3\over 16 \pi^3z^4}\ {({M\over z}+m)^2+{\bf k}_{\perp}^2\over
\Big( {\bf k}_{\perp}^2+{\lambda}_{R}^2({1\over z}){\Big)}^4}\ ,
\label{sa4f}\\
H_1^R(z,z^2{\bf k}_{\perp}^2)&=&a_R\
{{\bar N}_R^2(1-z)^3\over 16 \pi^3z^4}\ {({M\over z}+m)^2\over
\Big( {\bf k}_{\perp}^2+{\lambda}_{R}^2({1\over z}){\Big)}^4}\ ,
\label{sa5f}
\end{eqnarray}
where
\begin{equation}
{\lambda}_{R}^2({1\over z}) = (1-{1\over z}){\Lambda}^2 + {1\over z}M_R^2 - {1\over z}(1-{1\over z})M^2 \ ,
\label{ssa4a}
\end{equation}
with $M$, $m$ and $M_R$ are the hyperon, quark and diquark mass, respectively, and
\begin{eqnarray}
D_1^R(z)&=&
{{\bar N}_R^2\ z^2(1-z)^3\over 48\pi^2}\
{(M+mz)^2+{1\over 2}z^2{\lambda}_{R}^2({1\over z})\over \Big( z^2{\lambda}_{R}^2({1\over z}){\Big)}^3} \ ,
\label{transv5}\\
H_1^R(z)&=&a_R\
{{\bar N}_R^2\ z^2(1-z)^3\over 48\pi^2}\
{(M+mz)^2\over \Big( z^2{\lambda}_{R}^2({1\over z}){\Big)}^3} \ .
\label{transv6}\\
\end{eqnarray}
Here, we fix the normalization constant ${\bar N}_R$ by
\begin{equation}
\int_0^1 dz\ D_{1}^R(z)=1\ .
\label{ssa5}
\end{equation}
We normalized $D_{1}^R(z)$ in this way by considering that the number of hadrons
produced by a quark is determined by what is the number of that quark inside
the produced hadron.
The normalization constant ${\bar N}_R$ is cancelled in the ratio of Eq. (\ref{TP2})
since it is an overall multiplicative constant.

From the SU(6) wavefunctions written in Appendix A, we get the relations for $D_1$
given in Eq. (\ref{sa14}) for $\Lambda$ hyperon,
and the same relations are also satisfied for $H_1$.
The relations for other octet baryons can be obtained from
Eqs. (\ref{su12p}) to (\ref{su15}).
For example, Eq. (\ref{su12}) gives the relations for $\Sigma^+$:
\begin{eqnarray}
D_1^{u\to \Sigma^+}&=&2\ \Big( {3\over 4}D_1^s+{1\over 4}D_1^a\Big) \ ,
\qquad D_1^{d\to \Sigma^+}=0\ ,
\qquad D_1^{(s)\to \Sigma^+}=D_1^a
\ ,
\label{sigmaplusd1}\\
H_1^{u\to \Sigma^+}&=&2\ \Big( {3\over 4}H_1^s+{1\over 4}H_1^a\Big) \ ,
\qquad H_1^{d\to \Sigma^+}=0\ ,
\qquad H_1^{(s)\to \Sigma^+}=H_1^a
\ .
\label{sigmaplush1}
\end{eqnarray}

\begin{figure} 
\centering
\psfrag{xperp}[cc][cc]{$x_\perp$}
\begin{minipage}[t]{7.0cm}
\centering
\includegraphics[width=\textwidth]{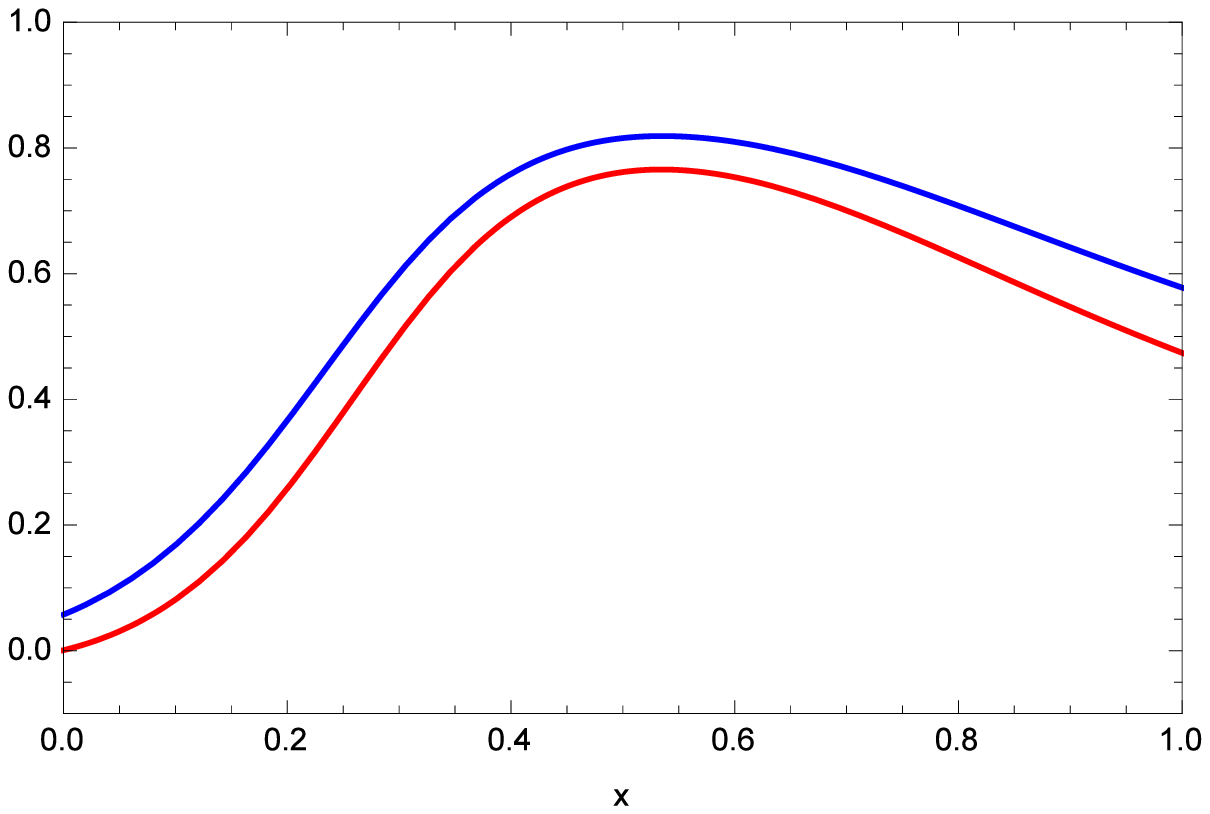}
(a)
\end{minipage}\hspace{0.5cm}
\begin{minipage}[t]{7.0cm}
\centering
\includegraphics[width=\textwidth]{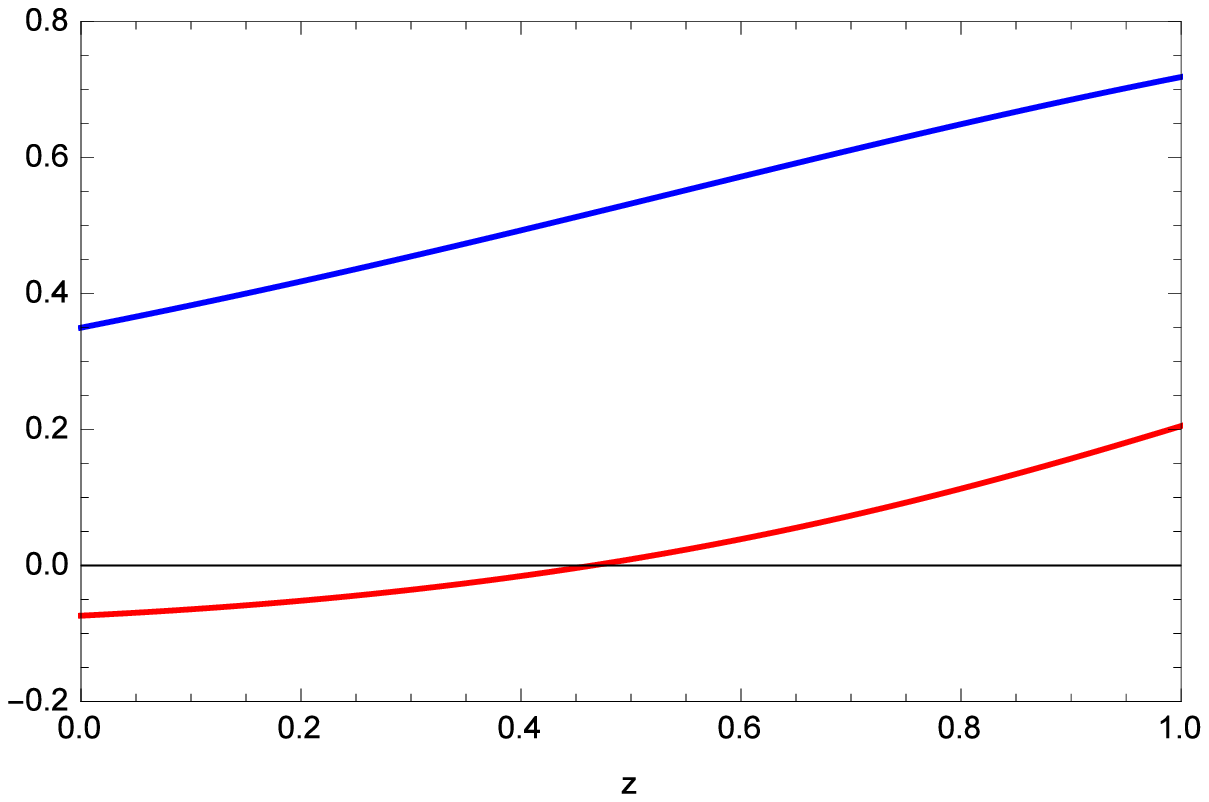}
(b)
\end{minipage}
\parbox{0.95\textwidth}{\caption{
(a) $F(x)$ of $\Lambda$ (lower line) and $F(x)$ of $\Sigma^+$ (upper line)
for the proton target.
(b) $G(z)$ of $\Lambda$ (lower line) and $G(z)$ of $\Sigma^+$ (upper line).
$F(x)$ and $G(z)$ are defined in Eq. (\ref{TP2aa}).
\label{f1h1udproton}}}
\end{figure}

\begin{figure}
\centering
\psfrag{xperp}[cc][cc]{$x_\perp$}
\begin{minipage}[t]{7.0cm}
\centering
\includegraphics[width=\textwidth]{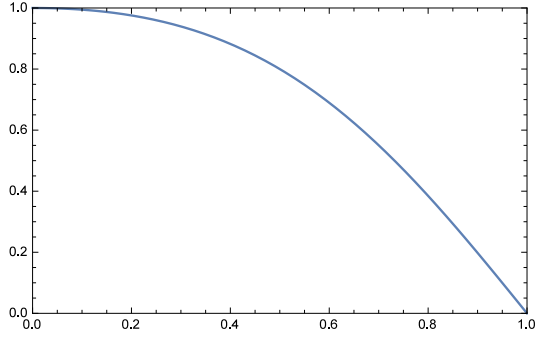}
\end{minipage}\hspace{0.5cm}
\parbox{0.95\textwidth}{\caption{
The $y$-dependent front factor $f(y)=(2(1-y))/(1+(1-y)^2)$ in (\ref{TP2aa}).
\label{frontfactor}}}
\end{figure}

\begingroup
\begin{table} 
\vspace*{1.2cm}
\label{tab1}
\begin{center}
\begin{tabular}{|c|c|c|c|c|}
\hline\hline
$h$ &$u$&$d$&$s$&$F^{N}(x)\ G^{h}(z)$\\
\hline\hline
\ \ \ $\Lambda$ \ \ \ & \ \ \ $({1\over 4}s+{3\over 4}a)$ \ \ \ &
\ \ \ $({1\over 4}s+{3\over 4}a)$ \ \ \  & \ \ \ $s$ \ \ \  &
\ \ \ \ \ $( {4h_{1u}+h_{1d}\over 4f_{1u}+f_{1d}}{)}^{N}\ ( {H_{1u}\over D_{1u}}{)}^{\Lambda}$ \ \ \ \ \ \\
\hline
\ \ \ $\Sigma^+$ \ \ \ & \ \ \ $(2)({3\over 4}s+{1\over 4}a)$ \ \ \ &
\ \ \ $0$ \ \ \  & \ \ \ $a$ \ \ \  &
\ \ \ \ \ $( {h_{1u}\over f_{1u}}{)}^{N}\ ( {H_{1u}\over D_{1u}}{)}^{\Sigma^+}$ \ \ \ \ \ \\
\hline
\ \ \ $\Sigma^0$ \ \ \ & \ \ \ $({3\over 4}s+{1\over 4}a)$ \ \ \ &
\ \ \ $({3\over 4}s+{1\over 4}a)$ \ \ \  & \ \ \ $a$ \ \ \  &
\ \ \ \ \ $( {4h_{1u}+h_{1d}\over 4f_{1u}+f_{1d}}{)}^{N}\ ( {H_{1u}\over D_{1u}}{)}^{\Sigma^0}$ \ \ \ \ \ \\
\hline
\ \ \ $\Sigma^-$ \ \ \ & \ \ \ $0$ \ \ \ &
\ \ \ $(2)({3\over 4}s+{1\over 4}a)$ \ \ \  & \ \ \ $a$ \ \ \  &
\ \ \ \ \ $( {h_{1d}\over f_{1d}}{)}^{N}\ ( {H_{1d}\over D_{1d}}{)}^{\Sigma^-}$ \ \ \ \ \ \\
\hline
\ \ \ $\Xi^0$ \ \ \ & \ \ \ $a$ \ \ \ &
\ \ \ $0$ \ \ \  & \ \ \ $(2)({3\over 4}s+{1\over 4}a)$ \ \ \  &
\ \ \ \ \ $( {h_{1u}\over f_{1u}}{)}^{N}\ ( {H_{1u}\over D_{1u}}{)}^{\Xi^0}$ \ \ \ \ \ \\
\hline
\ \ \ $\Xi^-$ \ \ \ & \ \ \ $0$ \ \ \ &
\ \ \ $a$ \ \ \  & \ \ \ $(2)({3\over 4}s+{1\over 4}a)$ \ \ \  &
\ \ \ \ \ $( {h_{1d}\over f_{1d}}{)}^{N}\ ( {H_{1d}\over D_{1d}}{)}^{\Xi^-}$ \ \ \ \ \ \\
\hline\hline
\end{tabular}
\end{center}
\vspace*{-0.5cm}
\caption{The scalar and axial-vector diquark contents in the octet baryons ($h$)
and the function $F(x)\, G(z)$ in (\ref{TP2aa}) for the nucleon ($N$) target.}
\end{table}
\endgroup

\begin{figure}
\centering
\psfrag{xperp}[cc][cc]{$x_\perp$}
\begin{minipage}[t]{7.4cm}
\centering
\includegraphics[width=\textwidth]{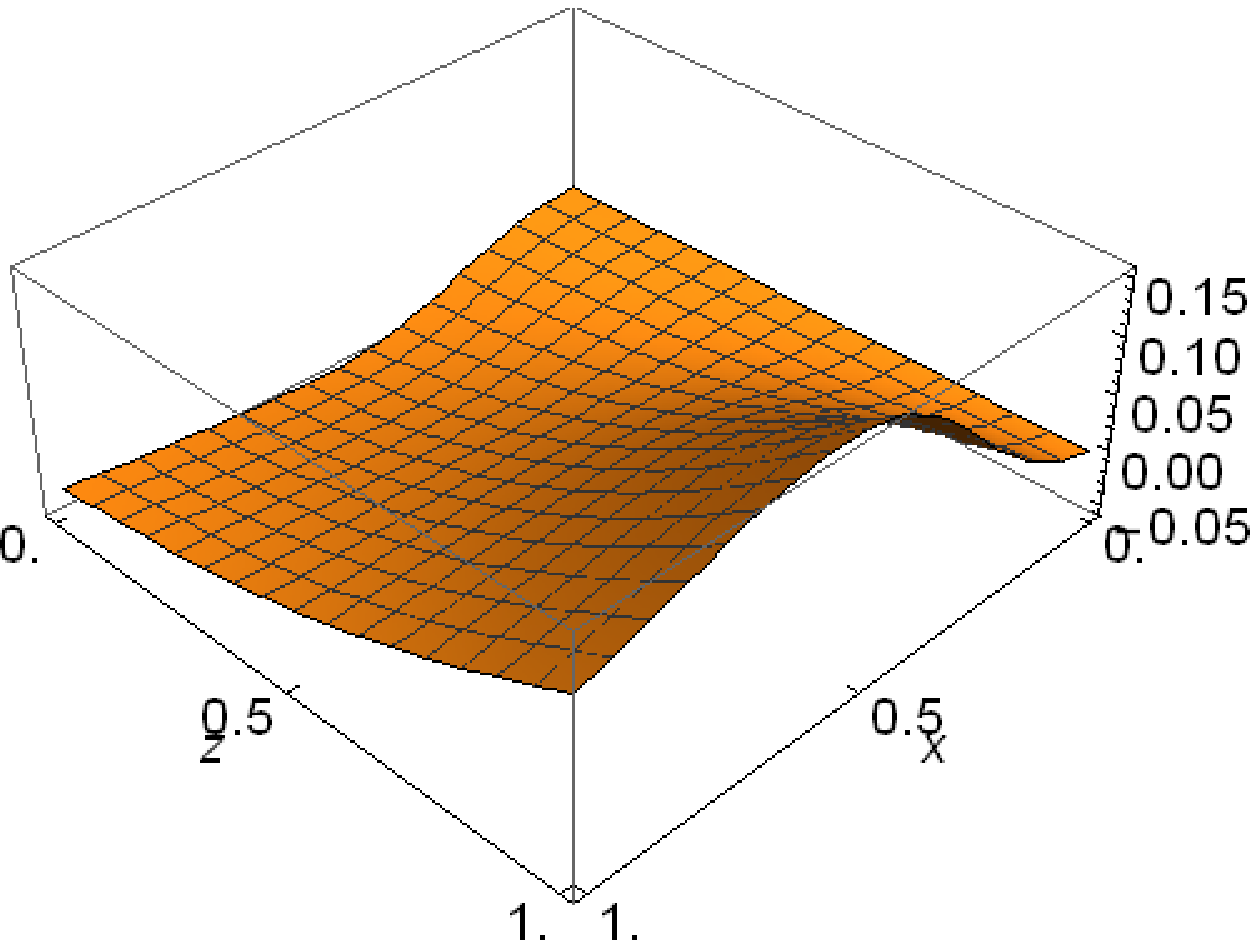}
(a) $\Lambda^0$
\end{minipage}\hspace{0.0cm}
\begin{minipage}[t]{7.4cm}
\centering
\includegraphics[width=\textwidth]{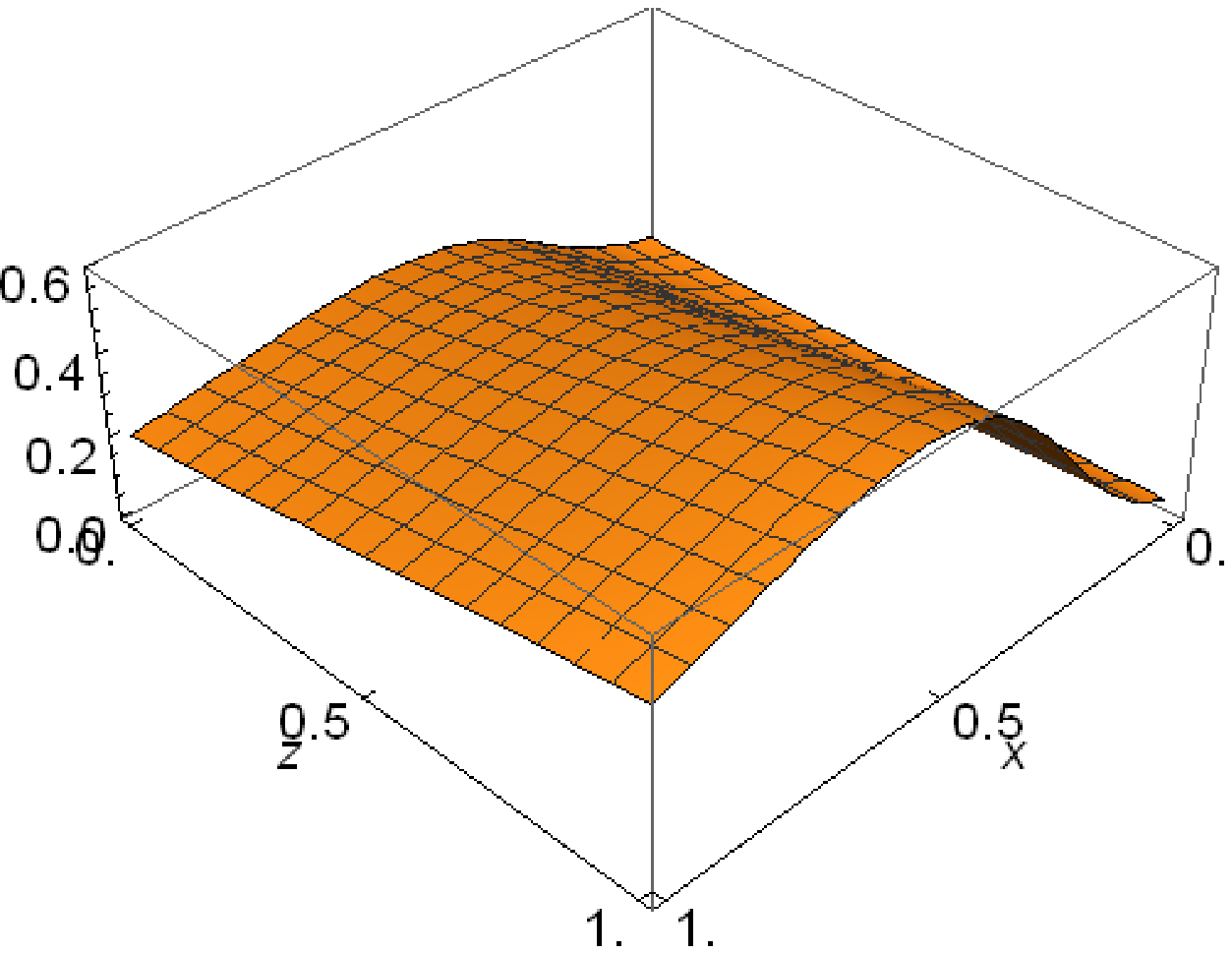}
(b) $\Sigma^+$
\end{minipage}\hspace{0.0cm}
\begin{minipage}[t]{7.4cm}
\centering
\includegraphics[width=\textwidth]{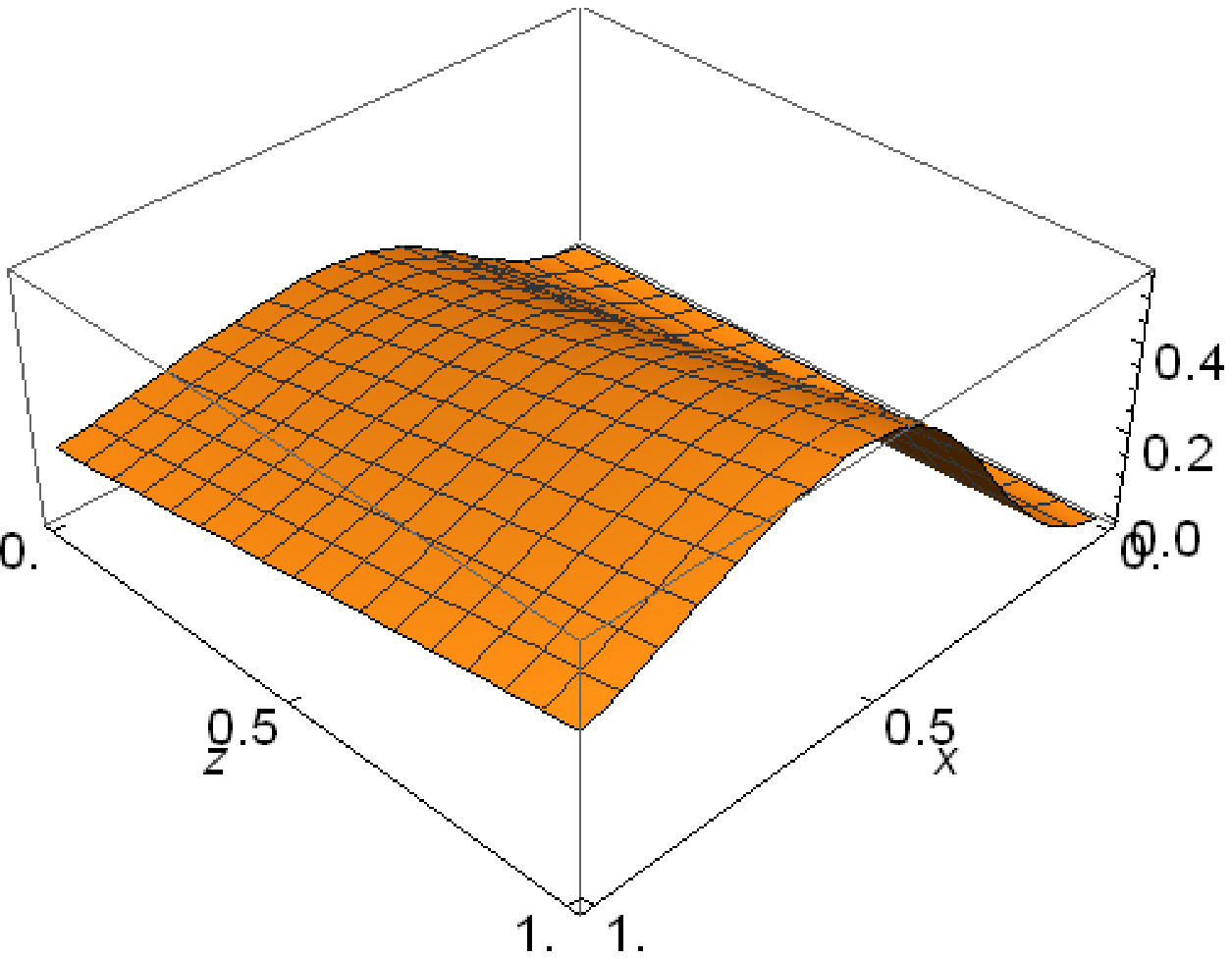}
(c) $\Sigma^0$
\end{minipage}
\begin{minipage}[t]{7.4cm}
\centering
\includegraphics[width=\textwidth]{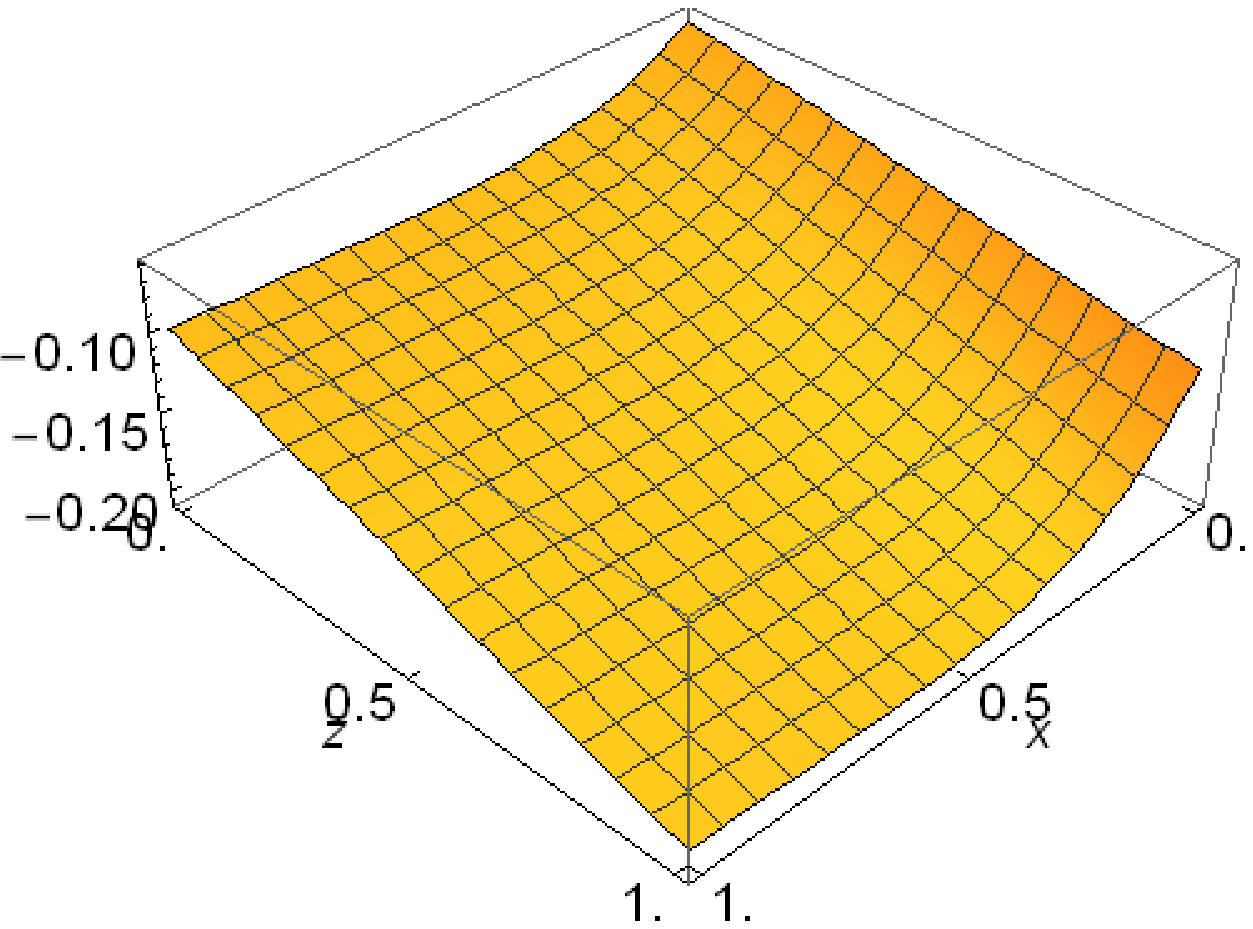}
(d) $\Sigma^-$
\end{minipage}\hspace{0.0cm}
\begin{minipage}[t]{7.4cm}
\centering
\includegraphics[width=\textwidth]{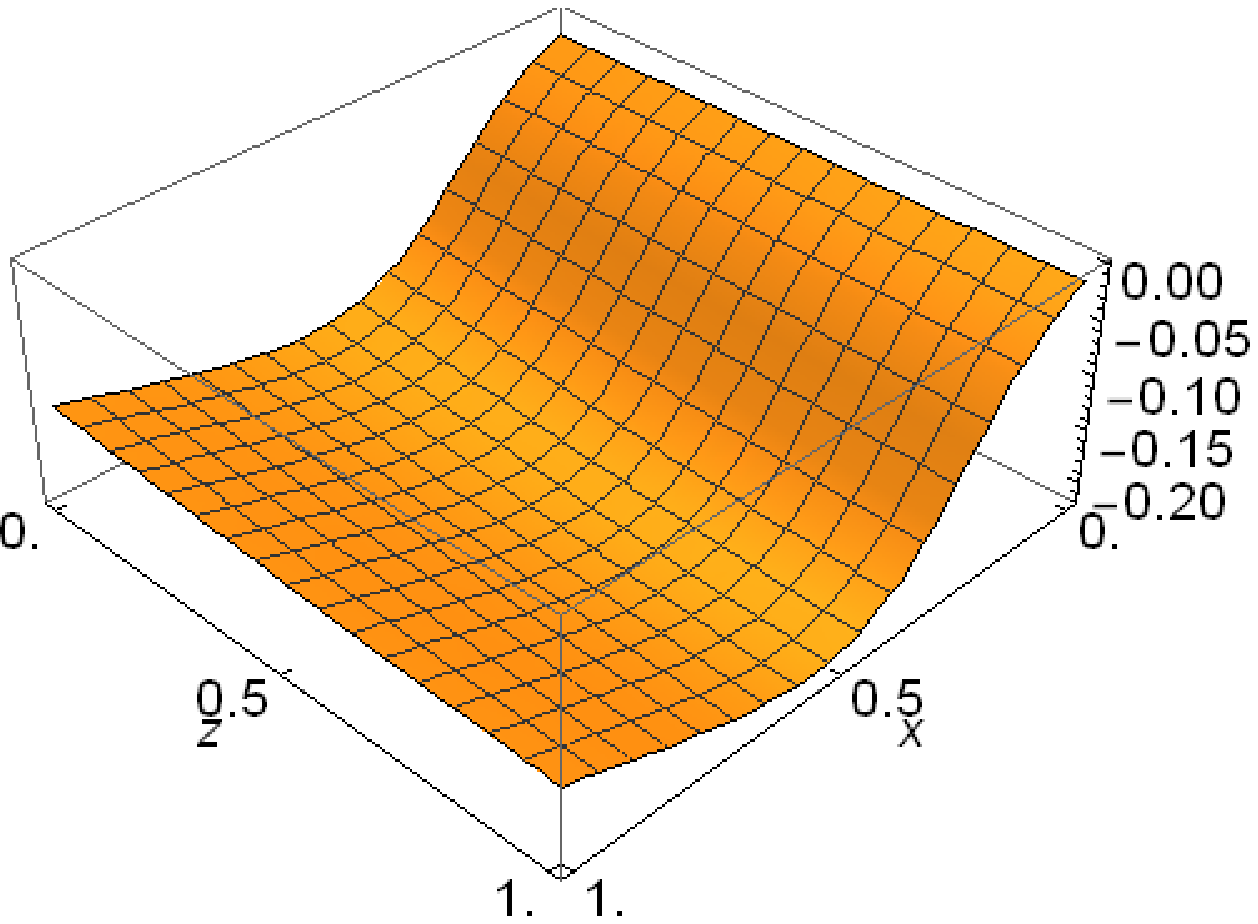}
(e) $\Xi^0$
\end{minipage}\hspace{0.0cm}
\begin{minipage}[t]{7.4cm}
\centering
\includegraphics[width=\textwidth]{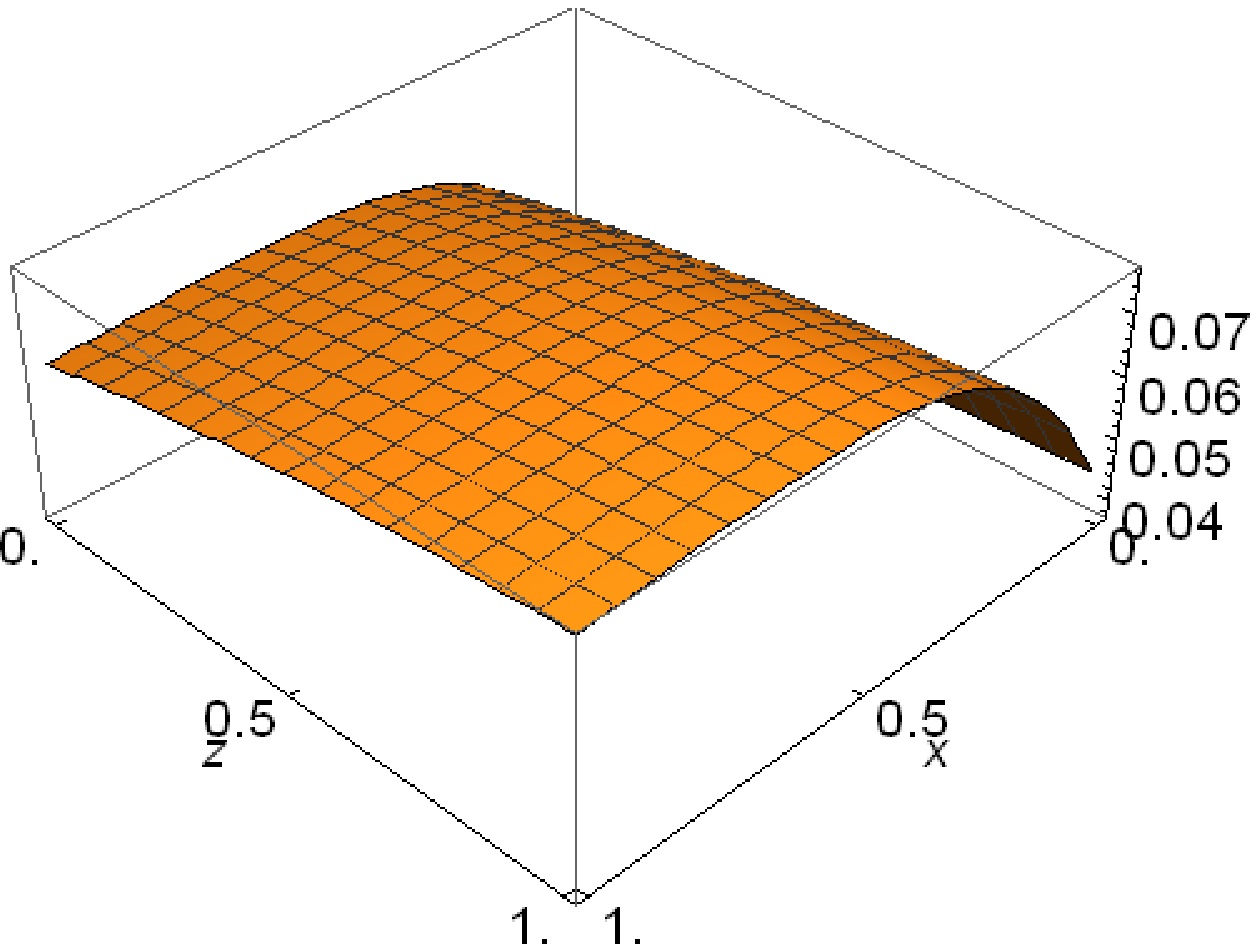}
(f) $\Xi^-$
\end{minipage}
\parbox{0.95\textwidth}{\caption{
$F(x)\, G(z)$ in (\ref{TP2aa})
for the hyperons produced from the proton target.
\label{hyperonsproton}}}
\end{figure}

\begin{figure}
\centering
\psfrag{xperp}[cc][cc]{$x_\perp$}
\begin{minipage}[t]{7.4cm}
\centering
\includegraphics[width=\textwidth]{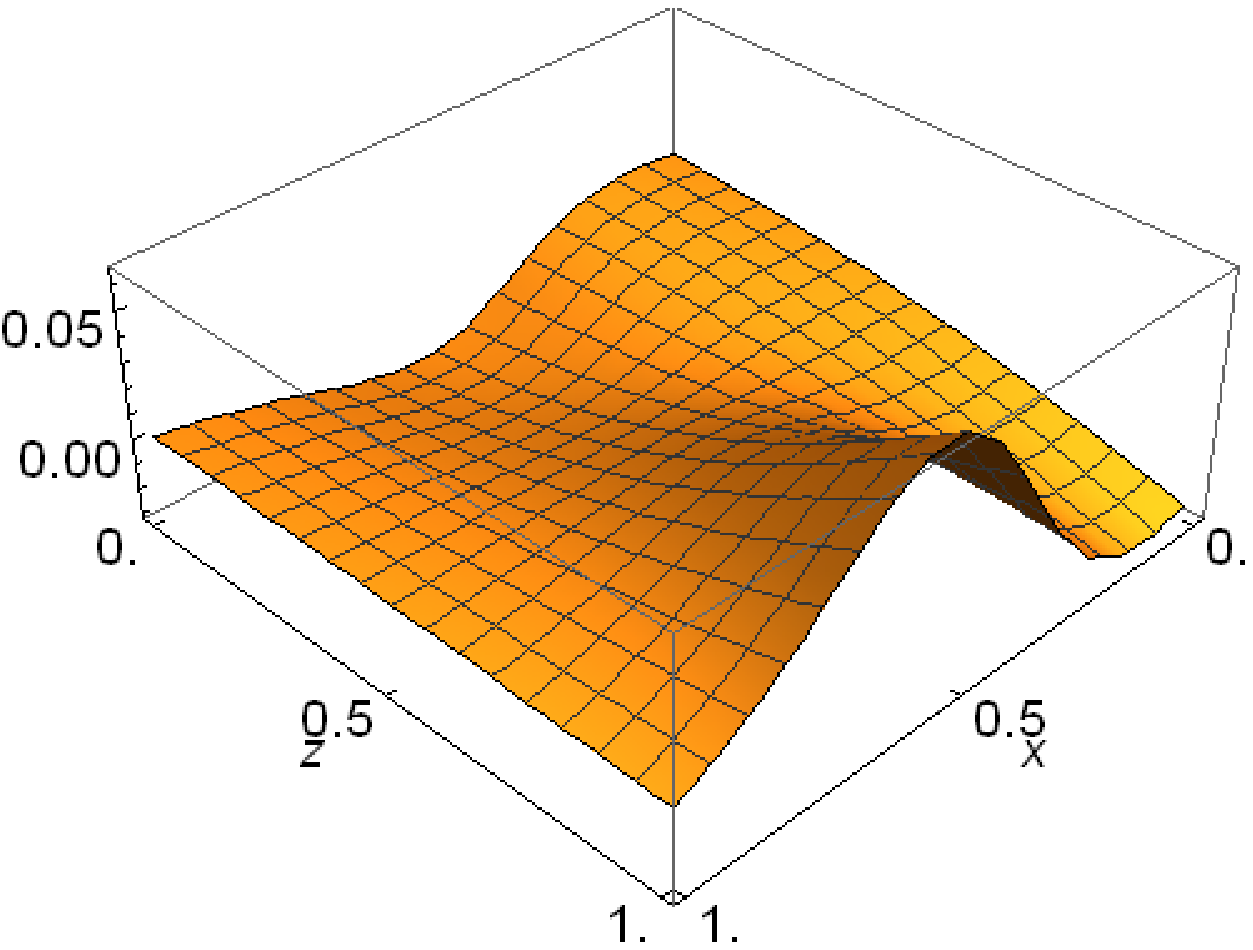}
(a) $\Lambda^0$
\end{minipage}\hspace{0.0cm}
\begin{minipage}[t]{7.4cm}
\centering
\includegraphics[width=\textwidth]{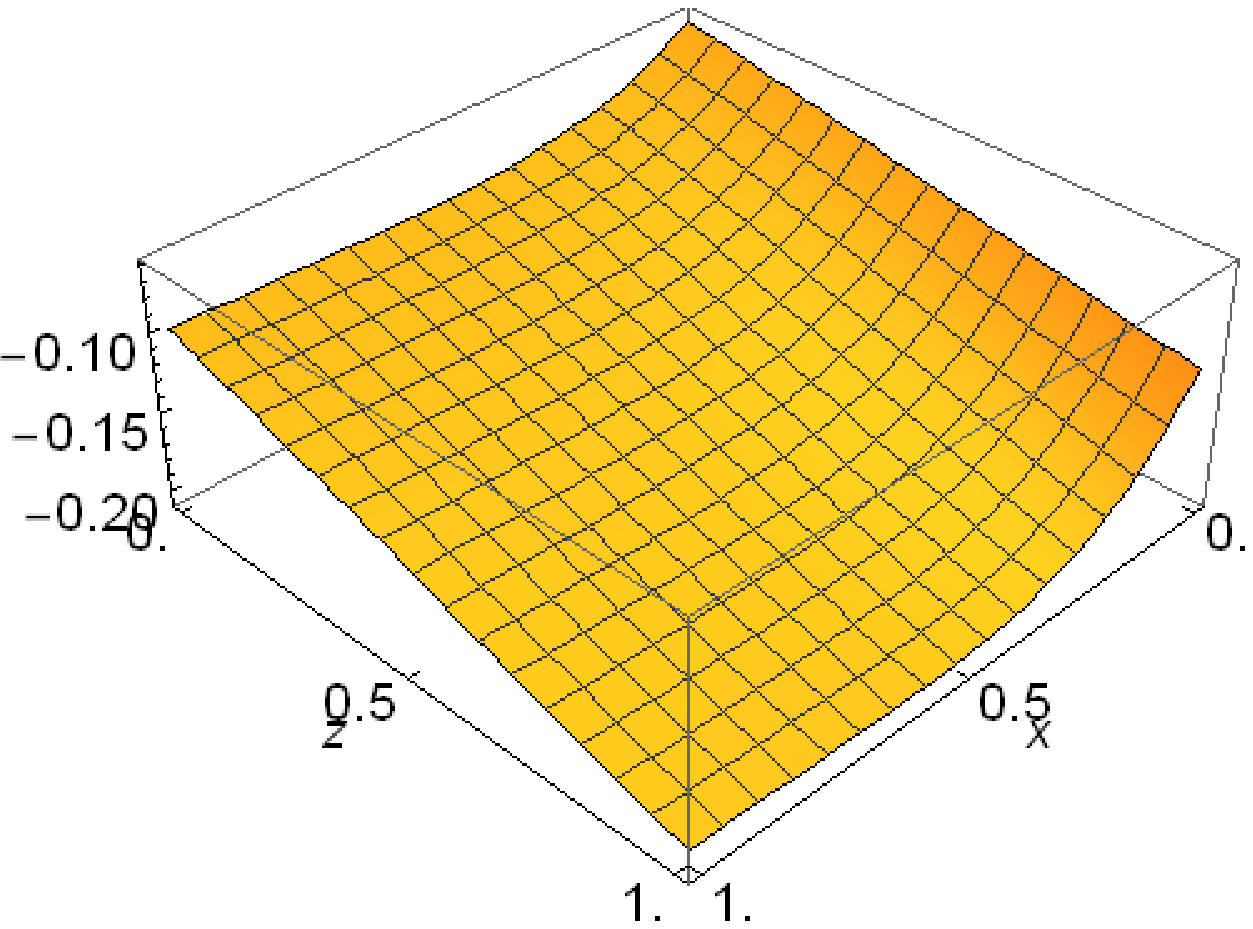}
(b) $\Sigma^+$
\end{minipage}\hspace{0.0cm}
\begin{minipage}[t]{7.4cm}
\centering
\includegraphics[width=\textwidth]{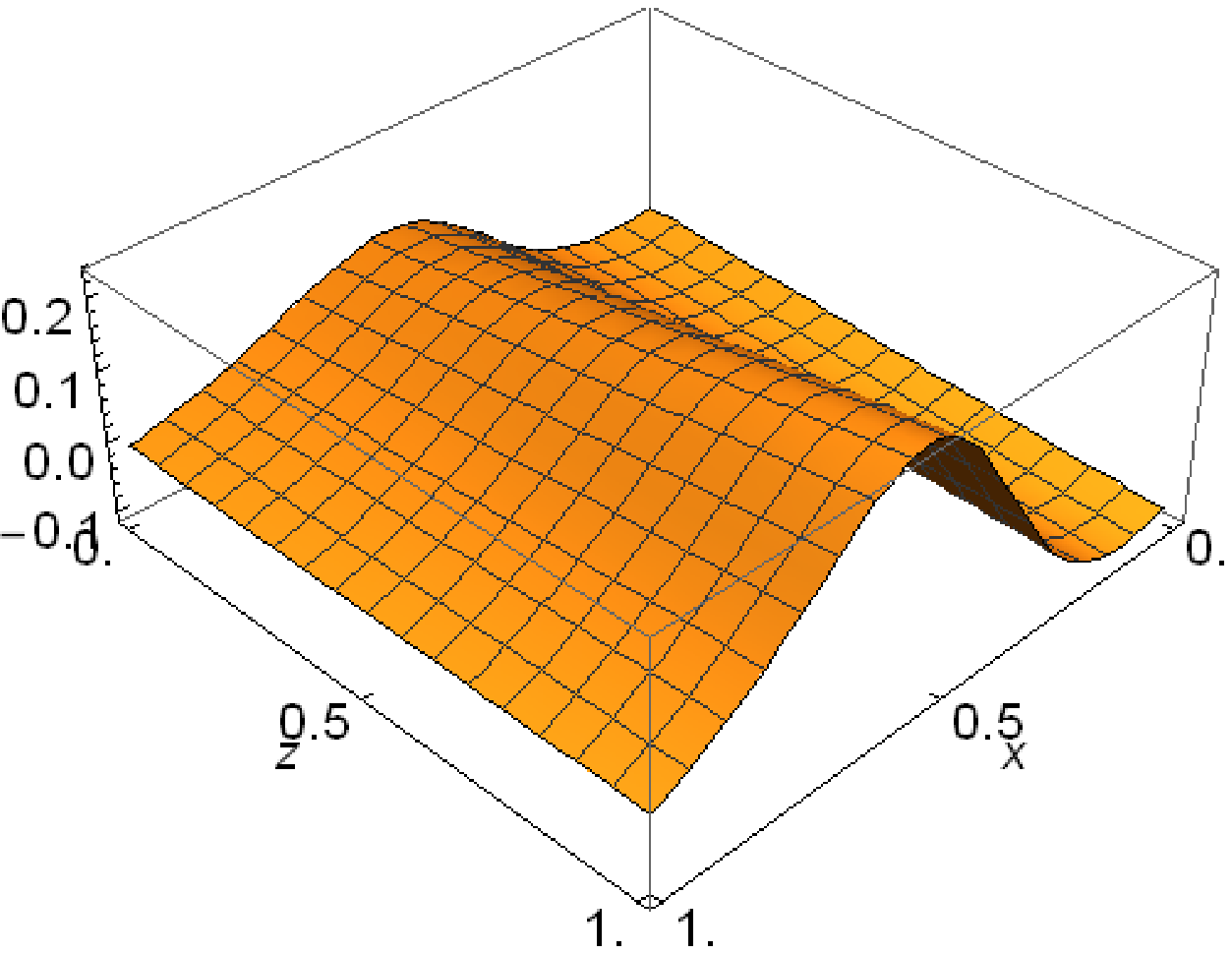}
(c) $\Sigma^0$
\end{minipage}
\begin{minipage}[t]{7.4cm}
\centering
\includegraphics[width=\textwidth]{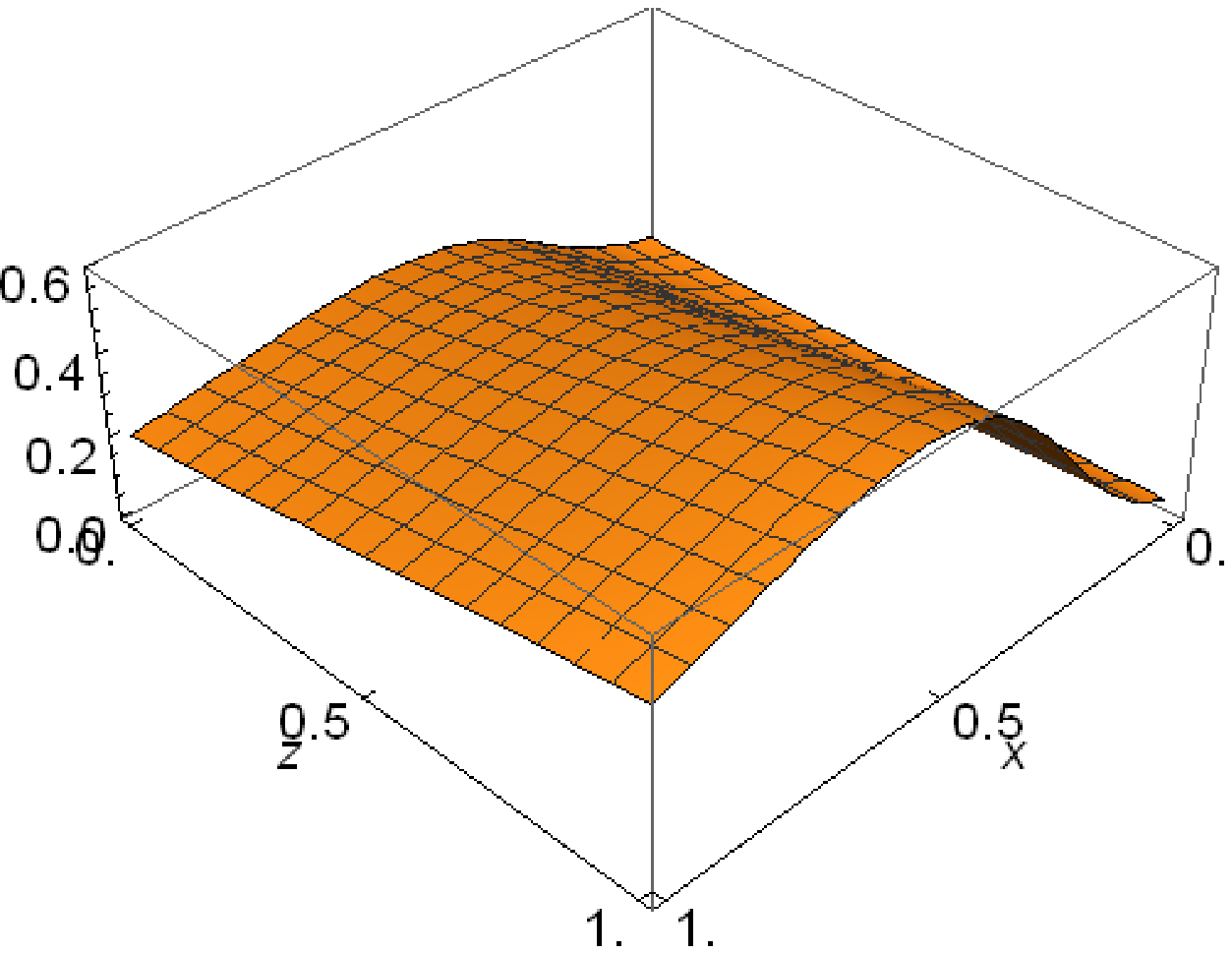}
(d) $\Sigma^-$
\end{minipage}\hspace{0.0cm}
\begin{minipage}[t]{7.4cm}
\centering
\includegraphics[width=\textwidth]{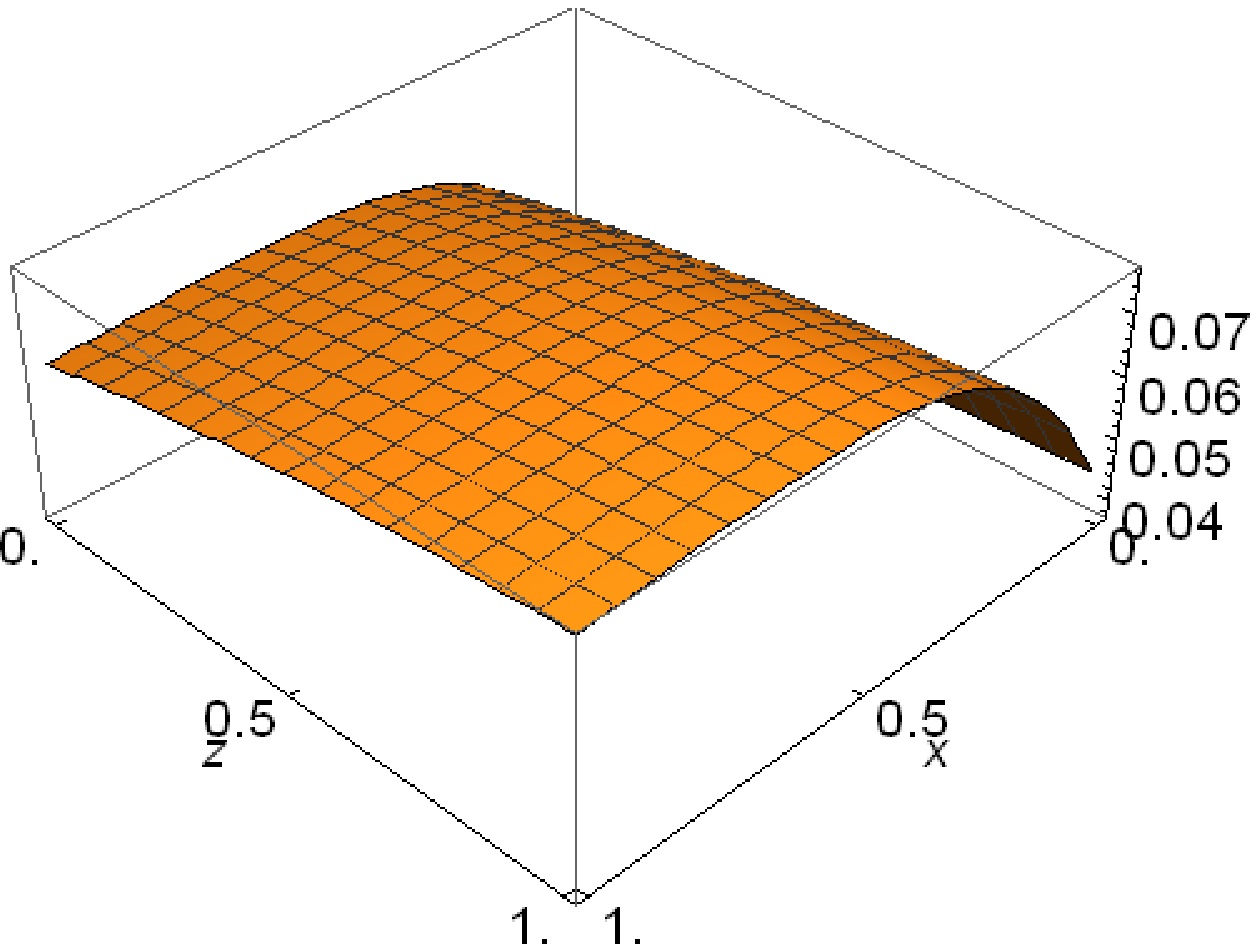}
(e) $\Xi^0$
\end{minipage}\hspace{0.0cm}
\begin{minipage}[t]{7.4cm}
\centering
\includegraphics[width=\textwidth]{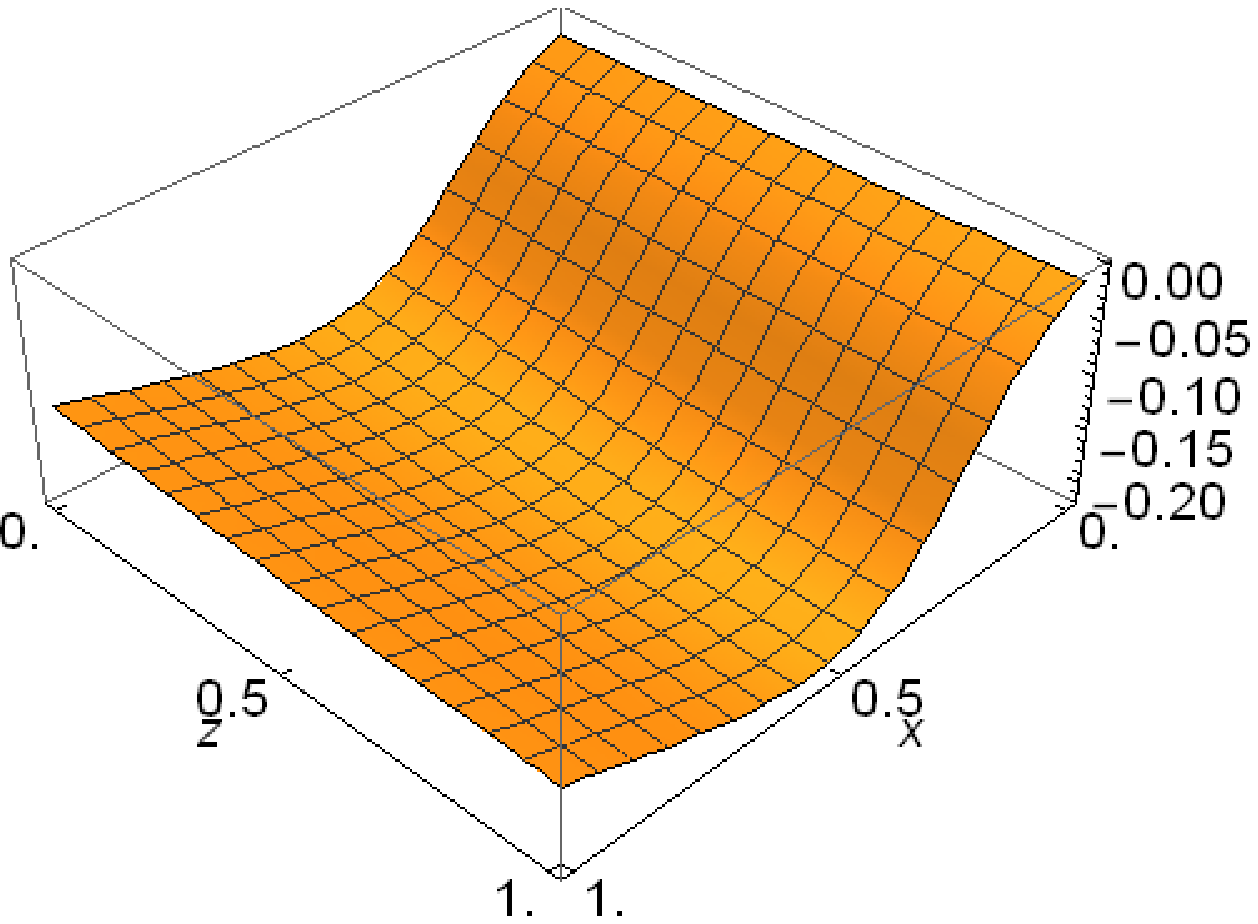}
(f) $\Xi^-$
\end{minipage}
\parbox{0.95\textwidth}{\caption{
$F(x)\, G(z)$ in (\ref{TP2aa})
for the hyperons produced from the neutron target.
\label{hyperonsneutron}}}
\end{figure}

\section{Polarizations of Hyperons}

Using the distribution functions of proton and the fragmentation functions of
hyperons given in section 3, we caculate the polarizations given in (\ref{TP2})
of the produced hyperons in the SIDIS process
of unpolarized lepton beam on transversely polarized proton target.

For the $\Lambda$ hyperon, we use $\Lambda =0.5$, $m=m_u=m_d=0.36$, $M_s=0.8$,
$M_a=1.0$ and $M=1.116$ ($\Lambda$ hyperon mass) in the unit of GeV.
Since $D_1^{u\to \Lambda}= D_1^{d\to \Lambda}$ and $H_1^{u\to \Lambda}= H_1^{d\to \Lambda}$
from (\ref{su11}), 
we find that Eq. (\ref{TP2}) can be written as
\begin{equation}
P_N={2(1-y)\over 1+(1-y)^2}\
{4h_{1u}(x) + h_{1d}(x) \over 4f_{1u}(x) + f_{1d}(x)}\
{H_{1u}(z) \over D_{1u}(z)}
\equiv
{2(1-y)\over 1+(1-y)^2}\ F(x)\ G(z)\ .
\label{TP2aa}
\end{equation}
Therefore, the $x$ and $z$ dependent part is factorized as a product of $F(x)$ and $G(z)$
for the $\Lambda$ production, and this factorization also happens for other five hyperons.
$F(x)$ and $G(z)$ for all hyperons are presented in the last column in Table 1.
However, this factorization does not happen for the proton and neutron productions as
presented in Appendix B.
The fragmentation functions are obtained from the formulas given in section 3.2.
$D_1^{u\to \Lambda}$ and $H_1^{u\to \Lambda}$ are presented in Fig. \ref{NormDH1uLambdaf},
and $D_1^{u\to \Sigma^+}$ and $H_1^{u\to \Sigma^+}$ in Fig. \ref{NormDH1uSigmaplusf}.
For the $\Lambda$ ($\Sigma^+$) production from the proton target, $F(x)$ and $G(z)$ are presented as
the lower (upper) line in Fig. \ref{f1h1udproton}.
The $y$-dependent front factor $f(y)=(2(1-y))/(1+(1-y)^2)$ in Eq. (\ref{TP2aa})
is drawn in Fig. \ref{frontfactor}.
The $x$ and $z$ dependent part $F(x)\, G(z)$ in Eq. (\ref{TP2aa}) for the $\Lambda$ ($\Sigma^+$)
production from the proton target is presented in Fig. \ref{hyperonsproton}(a)
(in Fig. \ref{hyperonsproton}(b)),
and for other hyperons in Fig. \ref{hyperonsproton}(c)-(f).
We find in Fig. \ref{hyperonsproton} that the polarization of the $\Lambda$ hyperon
is much smaller than the polarizations of $\Sigma^+$ and $\Sigma^0$,
the reason for which is that $H_1^{u\to \Lambda}$ is much smaller than $H_1^{u\to \Sigma^+}$
and $H_1^{u\to \Sigma^0}$.
(Figs. \ref{NormDH1uLambdaf}(b) and \ref{NormDH1uSigmaplusf}(b) show that
$H_1^{u\to \Lambda}$ is much smaller than $H_1^{u\to \Sigma^+}$.)
Therefore, when one tries to extract the transversity distribution $h_1(x)$
from the SIDIS process $lp^{\uparrow}\to lH^{\uparrow}X$ on the proton target,
investigating by measuring the polarizations of $\Sigma^+$ and $\Sigma^0$ is more
efficient than the polaization of $\Lambda$.
The $x$ and $z$ dependent parts $F(x)\, G(z)$ in the case of the neutron target
are presented in Fig. \ref{hyperonsneutron},
which shows that the polarizations of $\Sigma^-$ and $\Sigma^0$ are much larger
than that of $\Lambda$.
When we consider that the polarization of ${\Sigma}^0$ is measured throuth the daughter
$\Lambda$ polarization,
measuring the polarization of $\Sigma^+$ is most efficient for the proton target,
and $\Sigma^-$ for the neutron target.

\section{Conclusion}

We calculated with the SU(6) wavefunctions of the octet baryons and the spectator model
the transverse polarizations of the produced hyperons in the SIDIS
of unpolarized lepton beam on transversely polarized nucleon target,
since these polarizations provide a potential method for extracting the transversity
distribution $h_1(x)$ of the nucleon.
We find that the SU(6) wavefunctions imply that
$H_1(z)$ of the $\Lambda$ hyperon
is very small for both initial $u$ and $d$ quarks, and therefore
the polarization of the produced $\Lambda$ is very small for both transversely
polarized proton and neutron targets.
On the other hand, $H_1(z)$ of $\Sigma^+$ is large for initial $u$ quark and
$H_1(z)$ of $\Sigma^0$ is large for both initial $u$ and $d$ quarks, and
$H_1(z)$ of $\Sigma^-$ is large for initial $d$ quark.
Therefore, when one tries to extract the transversity distribution $h_1(x)$
from the SIDIS $lp^{\uparrow}\to lH^{\uparrow}X$,
it is expected that the polarizations of the $\Sigma^+$ and $\Sigma^0$ hyperons are large
for proton target and the polarizations of the $\Sigma^-$ and $\Sigma^0$ hyperons are large
for neutron target.
When we consider that the polarization of ${\Sigma}^0$ is measured throuth the daughter
$\Lambda$ polarization,
for extracting $h_1(x)$ from the SIDIS process $lp^{\uparrow}\to lH^{\uparrow}X$
it is most efficient to work by measuring the polarization of $\Sigma^+$ for the proton target,
and the polarization of $\Sigma^-$ for the neutron target.

Since the discovery of large $\Lambda$ polarization in $p+{\rm Be}\to {\Lambda}^0+X$ at
300 GeV \cite{Bunce:1976yb}, the polarizations of other hyperons were measured in
$p+{\rm Be}\to {\rm Hyperon}+X$ at 400 GeV: ${\Sigma}^+$ \cite{Wilkinson:1981jy},
${\Sigma}^-$ \cite{Deck:1983xu}, ${\Xi}^0$ \cite{Heller:1983ia} and ${\Xi}^-$ \cite{Rameika:1986rb}.
The polarization of ${\Sigma}^0$ was measured in $p+{\rm Be}\to {\Sigma}^0+X$ at
28.5 GeV throuth the daughter $\Lambda$ polarization \cite{Dukes:1987ys}.
The polarization was also measured for the $\Lambda$ hyperons which were produced
by unpolarized protons incident on hydrogen and deuterium targets \cite{Raychaudhuri:1980dx},
and a lot of experiments measuring the hyperon polarizations have been performed
in order to explore the nature of this single-spin asymmetry phenomenon.
Therefore, it is expected that it is possible to perform experiments which measure
the polarizations of all the six hyperons in the SIDIS process $lp^{\uparrow}\to l{\rm H}^{\uparrow}X$.
The results of these measurements would provide the information of the transversity of nucleon,
and also of the hyperon structures.

\section*{Acknowledgements}
This work was supported in part by the International Cooperation
Program of the KICOS (Korea Foundation for International Cooperation
of Science \& Technology).

\vfill\pagebreak

\section*{Appendix A: SU(6) wavefunctions of octet baryons}

\begin{eqnarray}
p^{\uparrow}&=&={1\over \sqrt{2}}(ud)_{0,0}u^{\uparrow}
\label{su62bp}\\
&&
+{1\over \sqrt{3}} \Big( \sqrt{2\over 3} (uu)_{1,1}d^{\downarrow} - \sqrt{1\over 3} (uu)_{1,0}d^{\uparrow} \Big)
-{1\over \sqrt{6}} \Big( \sqrt{2\over 3} (ud)_{1,1}u^{\downarrow} - \sqrt{1\over 3} (ud)_{1,0}u^{\uparrow} \Big)
\nonumber\\
n^{\uparrow}&=&={1\over \sqrt{2}}(ud)_{0,0}d^{\uparrow}
\label{su62bn}\\
&&
-{1\over \sqrt{3}} \Big( \sqrt{2\over 3} (dd)_{1,1}u^{\downarrow} - \sqrt{1\over 3} (dd)_{1,0}u^{\uparrow} \Big)
+{1\over \sqrt{6}} \Big( \sqrt{2\over 3} (ud)_{1,1}d^{\downarrow} - \sqrt{1\over 3} (ud)_{1,0}d^{\uparrow} \Big)
\nonumber\\
{\Lambda}^{\uparrow}&=&{1\over \sqrt{3}}(ud)_{0,0}s^{\uparrow}
+{1\over \sqrt{12}}(us)_{0,0}d^{\uparrow}-{1\over \sqrt{12}}(ds)_{0,0}u^{\uparrow}
\label{su61b}\\
&&
+{1\over 2}\Big( \sqrt{2\over 3} (us)_{1,1}d^{\downarrow} - \sqrt{1\over 3} (us)_{1,0}d^{\uparrow} \Big)
-{1\over 2}\Big( \sqrt{2\over 3} (ds)_{1,1}u^{\downarrow} - \sqrt{1\over 3} (ds)_{1,0}u^{\uparrow} \Big)
\nonumber\\
\Sigma^{+\uparrow}&=&{1\over \sqrt{2}}(us)_{0,0}u^{\uparrow}
\label{su62b}\\
&&
+{1\over \sqrt{3}} \Big( \sqrt{2\over 3} (uu)_{1,1}s^{\downarrow} - \sqrt{1\over 3} (uu)_{1,0}s^{\uparrow} \Big)
-{1\over \sqrt{6}} \Big( \sqrt{2\over 3} (us)_{1,1}u^{\downarrow} - \sqrt{1\over 3} (us)_{1,0}u^{\uparrow} \Big)
\nonumber\\
\Sigma^{0\uparrow}&=&{1\over 2}(us)_{0,0}d^{\uparrow}+{1\over 2}(ds)_{0,0}u^{\uparrow}
\label{su62bsig0}\\
&&
+{1\over \sqrt{3}} \Big( \sqrt{2\over 3} (ud)_{1,1}s^{\downarrow} - \sqrt{1\over 3} (ud)_{1,0}s^{\uparrow} \Big)
\nonumber\\
&&
-{1\over \sqrt{12}} \Big( \sqrt{2\over 3} (us)_{1,1}d^{\downarrow} - \sqrt{1\over 3} (us)_{1,0}d^{\uparrow} \Big)
-{1\over \sqrt{12}} \Big( \sqrt{2\over 3} (ds)_{1,1}u^{\downarrow} - \sqrt{1\over 3} (ds)_{1,0}u^{\uparrow} \Big)
\nonumber\\
\Sigma^{-\uparrow}&=&{1\over \sqrt{2}}(ds)_{0,0}d^{\uparrow}
\label{su63b}\\
&&
+{1\over \sqrt{3}} \Big( \sqrt{2\over 3} (dd)_{1,1}s^{\downarrow} - \sqrt{1\over 3} (dd)_{1,0}s^{\uparrow} \Big)
-{1\over \sqrt{6}} \Big( \sqrt{2\over 3} (ds)_{1,1}d^{\downarrow} - \sqrt{1\over 3} (ds)_{1,0}d^{\uparrow} \Big)
\nonumber\\
\Xi^{0\uparrow}&=&{1\over \sqrt{2}}(us)_{0,0}s^{\uparrow}
\label{su64b}\\
&&
-{1\over \sqrt{3}} \Big( \sqrt{2\over 3} (ss)_{1,1}u^{\downarrow} - \sqrt{1\over 3} (ss)_{1,0}u^{\uparrow} \Big)
+{1\over \sqrt{6}} \Big( \sqrt{2\over 3} (us)_{1,1}s^{\downarrow} - \sqrt{1\over 3} (us)_{1,0}s^{\uparrow} \Big)
\nonumber\\
\Xi^{-\uparrow}&=&{1\over \sqrt{2}}(ds)_{0,0}s^{\uparrow}
\label{su65b}\\
&&
-{1\over \sqrt{3}} \Big( \sqrt{2\over 3} (ss)_{1,1}d^{\downarrow} - \sqrt{1\over 3} (ss)_{1,0}d^{\uparrow} \Big)
+{1\over \sqrt{6}} \Big( \sqrt{2\over 3} (ds)_{1,1}s^{\downarrow} - \sqrt{1\over 3} (ds)_{1,0}s^{\uparrow} \Big)
\ .
\nonumber
\end{eqnarray}
\\

The $S^z=+{1\over 2}$ state of proton is given in (\ref{su62bp}) and
the $S^z=-{1\over 2}$ state is given by
\begin{eqnarray}
p^{\downarrow}&=&={1\over \sqrt{2}}(ud)_{0,0}u^{\downarrow}
\label{su62bpdown}\\
&&
+{1\over \sqrt{3}} \Big( - \sqrt{2\over 3} (uu)_{1,-1}d^{\uparrow} + \sqrt{1\over 3} (uu)_{1,0}d^{\downarrow} \Big)
-{1\over \sqrt{6}} \Big( - \sqrt{2\over 3} (ud)_{1,-1}u^{\uparrow} + \sqrt{1\over 3} (ud)_{1,0}u^{\downarrow} \Big)
\ .
\nonumber
\end{eqnarray}
That is, $(ud)_{0,0}u^{\uparrow}$,
$\Big( \sqrt{2\over 3} (uu)_{1,1}d^{\downarrow} - \sqrt{1\over 3} (uu)_{1,0}d^{\uparrow} \Big)$
and $\Big( \sqrt{2\over 3} (ud)_{1,1}u^{\downarrow} - \sqrt{1\over 3} (ud)_{1,0}u^{\uparrow} \Big)$
in (\ref{su62bp}) are replaced by
$(ud)_{0,0}u^{\downarrow}$,
$\Big( - \sqrt{2\over 3} (uu)_{1,-1}d^{\uparrow} + \sqrt{1\over 3} (uu)_{1,0}d^{\downarrow} \Big)$
and $\Big( - \sqrt{2\over 3} (ud)_{1,-1}u^{\uparrow} + \sqrt{1\over 3} (ud)_{1,0}u^{\downarrow} \Big)$
in (\ref{su62bpdown}). The $S^z=-{1\over 2}$ state of other octet baryons can be obtained in the same way.

\section*{Appendix B: Polarizarions of produced proton and neutron}

When hyperons are produced from the nucleon target, the $x$ and $z$ dependent
part in Eq. (\ref{TP2}) for $P_N$ is
factorized as $F(x)\, G(z)$ as shown in Eq. (\ref{TP2aa}).
However, when nucleons are produced, this factorization does not happen
and we should use Eq. (\ref{TP2}) as written in the last column in Table 2.
We have the following relations for $D_1$ of proton and neutron:
\begin{eqnarray}
D_1^{u\to p}&=&2\ \Big( {3\over 4}D_1^s+{1\over 4}D_1^a\Big) \ ,
\qquad D_1^{d\to p}=D_1^a\ ,
\label{protonD1}\\
D_1^{u\to n}&=&D_1^a\ ,
\qquad D_1^{d\to n}=2\ \Big( {3\over 4}D_1^s+{1\over 4}D_1^a\Big) \ ,
\label{protonD1}
\end{eqnarray}
and the same relations are also satisfied for $H_1$ of proton and neutron.

\begingroup
\begin{table} [b]
\vspace*{1.2cm}
\label{tablepn}
\begin{center}
\begin{tabular}{|c|c|c|c|c|}
\hline\hline
$h$ &$u$&$d$&$s$&$F^{Nh}(x,z)$\\
\hline\hline
\ \ \ $p$ \ \ \ & \ \ \ $(2)({3\over 4}s+{1\over 4}a)$ \ \ \ &
\ \ \ $a$ \ \ \  & \ \ \ $0$ \ \ \ &
\ \ \ \ \
${4h_{1u}^{N}(x)H_{1u}^{p}(z)+h_{1d}^{N}(x)H_{1d}^{p}(z)\over 4f_{1u}^{N}(x)D_{1u}^{p}(z)+f_{1d}^{N}(x)D_{1d}^{p}(z)}$
\ \ \ \ \  \\
\hline
\ \ \ $n$ \ \ \ & \ \ \ $a$ \ \ \ &
\ \ \ $(2)({3\over 4}s+{1\over 4}a)$ \ \ \  & \ \ \ $0$ \ \ \  &
\ \ \ \ \
${4h_{1u}^{N}(x)H_{1u}^{n}(z)+h_{1d}^{N}(x)H_{1d}^{n}(z)\over 4f_{1u}^{N}(x)D_{1u}^{n}(z)+f_{1d}^{N}(x)D_{1d}^{n}(z)}$
\ \ \ \ \  \\
\hline\hline
\end{tabular}
\end{center}
\vspace*{-0.5cm}
\caption{The scalar and axial-vector diquark contents in the nucleons ($h$)
and the function $F^{Nh}(x,z)$ is the $x$ and $z$ dependent part in (\ref{TP2})
for the nucleon ($N$) target.}
\end{table}
\endgroup

\vfill\pagebreak

\end{document}